# Energy of natural interface modes

**Illarion Dorofeyev**

Institute for Physics of Microstructures, Russian Academy of Sciences,

603950, GSP-105 Nizhny Novgorod, Russia

**Abstract**

The paper is devoted to the thermodynamics of normal surface electromagnetic fields within a nonuniform dispersive and absorptive system. This system is formed by vacuum and lossy medium separated by a plane interface. As a medium, we used dielectric and metal samples characterized by local and nonlocal optical properties. Thermodynamic properties of surface eigenmodes of plane interfaces are discussed. Various definitions of density of states and spectral characteristics of surface polaritons in equilibrium at the interface formed by vacuum and lossy medium are described and discussed. All formulas for thermodynamic functions are represented in terms of density of states. The generalized density of states is calculated based on the Barash-Ginzburg theory and dispersion relations for the surface states in different approaches. It is exemplified that different definitions of the density of states are identical in the case of dissipationless materials. The spectral functions and integrated over all frequencies thermodynamic characteristics and their temperature dependences are demonstrated.

**Keywords**: energy, interface, polaritons, density of states, dispersion equation

**PACS numbers:** 05.40.-a, 68.35.Md, 05.70.Np, 63.20.Pw

**Corresponding author:** Illarion Dorofeyev

Institute for Physics of Microstructures Russian Academy of Sciences,

603950 Nyzhny Novgorod, GSP-105, Russia

Phone:+7 910 7934517  Fax:+7 8312 675553

E-mail1: Illarion1955@mail.ru

E-mail2: dorof@ipm.sci-nnov.ru

# 1. Introduction

A system in thermodynamic equilibrium is fully characterized by its thermodynamic functions [1]. All these functions can be expressed in different ways; namely, in terms of the partition function, by using the density of states or local density of states, by using the Green's functions, or by using some auxiliary functions, which directly associated with the density of states and Green's functions [2]. As usual, thermodynamic quantities are well-known for bulk (infinite) systems. The related example that we want to mention here is the energy, free energy, entropy and thermal capacity of the blackbody photons in a cavity [1]. In general, natural and manufactured compound structures consist of volume and surface parts separated by transition layers that connect them. Here we consider a system of thermally stimulated electromagnetic fields nearby interfaces of adjoined materials separated by plane interfaces. Physics of thermal electromagnetic fields is well developed [3-16], but a problem of calculation of the thermal electromagnetic fields within compound structures is quite laborious, even for the case of lossless materials. Density of energy and spectral composition of the fluctuating fields are very important quantities. Usually, these properties are calculated in a vacuum space out of heated bodies. The matter is that the definition of the thermodynamic characteristics within dispersive and absorptive medium takes special attention, because the energy balance relation following from the Maxwell's equations contains the electromagnetic energy coupled with the dissipated heat in a system, and in general it is impossible to separate the two parts in an unambiguous way. But, the dissipated heat vanishes in average in case of thermal equilibrium. That is why; the electromagnetic energy within dispersive and absorbing medium in equilibrium can be defined well. The problem of defining and calculating an electromagnetic energy density for the classical EM field in dispersive and absorbing media was investigated in detail by Barash and Ginzburg [17, 18].

Our paper aims at establishing the correspondence between the Barash-Ginzburg approach in calculations of thermodynamic properties of the fluctuating electromagnetic fields within spatially nonuniform dispersive and dissipative systems and the traditional textbook definition of energy via the density of states of eigenmodes of spatially nonuniform dissipationless systems [19,20]. Another goal is to calculate these thermodynamic properties within a system of contacting bodies with different local and nonlocal optical properties using the general approach.

## 2. General equations

A general method of calculation of free energy and energy of the electromagnetic fields within spatially inhomogeneous and dissipative materials was developed in [17, 18]. As the result the expression, for instance, for the free energy can be represented as follows

$$F(T) = -\int_0^\infty \rho(\omega) k_B T \, \ell n[Z_0(\omega,T)] d\omega, \qquad (1)$$

where $Z_0(\omega,T)$ is the partition function,

$$Z_0(\omega,T) = \sum_{n=0}^\infty Exp[-E_n(\omega)/k_B T] = \frac{Exp(-\hbar\omega/2k_B T)}{1 - Exp(-\hbar\omega/k_B T)}, \qquad (2)$$

where $E_n(\omega) = \hbar\omega(n+1/2)$ are the energy levels of a free harmonic oscillator.

From Eq. (1) it is evident that the energy can be obtained by using the known relation [1]

$$U(T) = -T^2 \frac{\partial}{\partial T}\left(\frac{F(T)}{T}\right)_V = \int_0^\infty \rho(\omega) U(\omega,T) d\omega, \qquad (3)$$

where $U(\omega,T) = k_B T^2 \frac{\partial \ell n[Z(\omega,T)]}{\partial T} = \Theta(\omega,T)$, where $\Theta(\omega,T) = (\hbar\omega/2) cth(\hbar\omega/2k_B T)$ is the mean energy of a free oscillator. The entropy and heat capacity are obtained from Eqs. (1) and (2) by differentiation with respect to temperature [1]

$$S(T) = -\left(\partial F(T)/\partial T\right)_V, \quad C_V(T) = \left(\partial U(T)/\partial T\right)_V. \qquad (4)$$

For bosonic systems the chemical potential is equal to zero. That is why the free energy is equal to the thermodynamic potential $F(\omega,T) = \Omega(\omega,T) = -P(\omega,T)V$, where $P$ is the pressure and $V$ is the volume of a system under consideration.

The key quantity of the Barash-Ginzburg theory [17, 18], determining each of thermodynamic functions, is the integrand $\rho(\omega)$ in Eq. (1), given by the expression

$$\rho(\omega) = -\frac{1}{\pi} Im \left\{ \int_{\Omega_\lambda} \rho(\beta) d\beta \frac{\partial}{\partial \omega} \ell n D(\beta,\omega) \right\}, \qquad (5)$$

where $\beta$ is the variable connected with integrals of motion, in particular, the wavenumber components in a homogeneous infinite media with $\rho(\beta) = L^3/(2\pi)^3$ and $d\beta = k^2 dk \sin\vartheta d\vartheta d\phi$, $D = \prod_i D^{(i)}$, where $D^{(i)}$ is the dispersion equation, "$i$" means the branch of the natural modes of the system under study. For example [21], in an infinite homogeneous and dispersive media it is known two transverse eigenmodes of two independent polarizations, satisfying the dispersion equation $D^{(tr)} = k^2 - (\omega/c)^2 \varepsilon^{tr}(\omega,k)$ and one longitudinal eigenmode from the equation $D^{(\ell)} = \varepsilon^{\ell}(\omega,k)$. An integration in Eq.(5) is done over the volume $\Omega_\lambda$, separating the longwavelength part of the electromagnetic fields [17,18].

It should be noted that the density of states in a two-dimensional system can be calculated by analogy with a three-dimensional case usually employed in theory of solid states [19, 20]. This quantity can be written out as follows [21, 22]

$$\mathcal{D}(\omega) = \frac{S}{(2\pi)^2} \int \frac{d\ell_k}{|\nabla_{\vec{k}}\omega|}, \qquad (6)$$

where $S$ is the surface of sample under study, $d\ell_k$ is the differential element of a constant-frequency line $k_{SP}(\omega)$ in a two-dimensional wave-vector space, $\nabla_{\vec{k}}\omega$ is the group velocity of the surface polariton. In a two-dimensional wave-vector space the value $(2\pi)^2/S$ corresponds to one allowed state of $k_{SP}$.

Strictly speaking the definition of DOS in Eq.(6) means negligibly small absorptions in a system of interest ($\nu, \gamma \to 0$), where $\nu$ and $\gamma$ are the electron relaxation frequency and anharmonic decay constant in the Drude and oscillatory model of the dielectric function, correspondingly

$$\varepsilon(\omega) = 1 - \frac{\omega_P^2}{\omega(\omega + i\nu)}, \qquad (7)$$

and

$$\varepsilon(\omega) = \varepsilon_\infty \left[ 1 + \frac{(\omega_{LO}^2 - \omega_{TO}^2)}{\omega_{TO}^2 - \omega^2 - i\gamma\omega} \right], \qquad (8)$$

where $\omega_P$ is the plasma frequency, $\varepsilon_0, \varepsilon_\infty$ are the respective dielectric constants at low and high frequencies, $\omega_{TO}$ is the frequency of the transverse optical phonon. Here we mention the frequency of the quasistatic surface polariton $\omega_{QP}$, which corresponds to the nonradiative Coulomb surface polariton as a root of the dispersion equation $\text{Re}\{\varepsilon(\omega)\}=-1$. For Eqs.(7), (8) $\omega_{QP} = \omega_P/\sqrt{2}$ and $\omega_{QP} = \omega_{TO}\sqrt{(\varepsilon_0+1)/(\varepsilon_\infty+1)}$, correspondingly for dissipationless systems ($\nu = 0, \gamma = 0$).

We consider an isotropic material that is why the constant-frequency lines in a wavenumber space are the concentric circles in this case. The group velocity $\nabla_{\vec{k}}\omega$ is a constant along the constant-frequency line and $|\nabla_{\vec{k}}\omega| = |d\omega/dk_{SP}|$. The circle length is $\ell_k = 2\pi |k_{SP}|$. Taking into account these considerations, we can write from Eq.(6) the number of surface modes per unit frequency interval per unit surface

$$D(\omega) = \frac{\mathcal{D}(\omega)}{S} = \frac{|k_{SP}|}{2\pi}\left|\frac{dk_{SP}}{d\omega}\right|. \qquad (9)$$

Taking into account the finite dissipation rates we kept in Eq.(9) the absolute value of the surface polaritons wave number $k_{SP}(\omega) = p(\omega) + i\alpha(\omega)$.

A theory of surface electromagnetic waves is known from many textbooks and reviews, see for instance [23-27]. Surface polaritons at a vacuum–matter plane interface are characterized by the specific dispersion relation and validity condition as follows

$$k_{SP}^2 = \frac{\omega^2}{c^2}\frac{\varepsilon(\omega)}{\varepsilon(\omega)+1}, \qquad \mathrm{Re}\{\varepsilon(\omega)\} < -1, \qquad (10)$$

where $\varepsilon(\omega) = \varepsilon'(\omega) + i\varepsilon''(\omega)$ is an isotropic, frequency-dependent, complex dielectric function. Following to [23] we consider two cases: of the complex wavenumber of the surface polariton and a pure real frequency and of the complex frequency and a pure real wavenumber.

In case of a pure real frequency we separate the real and imaginary parts of the complex wavenumber in Eq.(10)

$$k_{SP}(\omega) = k'_{SP}(\omega) + ik''_{SP}(\omega) \equiv p(\omega) + i\alpha(\omega), \qquad (11)$$

which are equal to

$$p^2(\omega) = \frac{\omega^2}{2c^2}\left[\frac{(\varepsilon' + \varepsilon'^2 + \varepsilon''^2) + \sqrt{(\varepsilon' + \varepsilon'^2 + \varepsilon''^2)^2 + \varepsilon''^2}}{(\varepsilon'+1)^2 + \varepsilon''^2}\right], \qquad (12)$$

$$\alpha^2(\omega) = \frac{\omega^2}{2c^2}\left[\frac{\sqrt{(\varepsilon' + \varepsilon'^2 + \varepsilon''^2)^2 + \varepsilon''^2} - (\varepsilon' + \varepsilon'^2 + \varepsilon''^2)}{(\varepsilon'+1)^2 + \varepsilon''^2}\right], \qquad (13)$$

where we selected the roots displaying correct results for $k'^2_{SP} = (\omega^2/c^2)\varepsilon'(\omega)/(\varepsilon'(\omega)+1)$ and $k''_{SP} = 0$ in case of transition to the transparent medium $\varepsilon''(\omega) \to 0$.

By definition, the real and the imaginary parts in Eqs.(12) and (13) of a wavenumber are expressed via the wavelength $\lambda(\omega)$ and the propagation length $L(\omega)$ of surface polaritons

$$k_{SP}(\omega) = \frac{2\pi}{\lambda(\omega)} + i\frac{1}{L(\omega)} \equiv p(\omega) + i\alpha(\omega), \qquad (14)$$

in order to obtain an appropriate solution in the form $\exp\{i[k_{SP}(\omega)r_\| - \omega t]\} = \exp\{i[p(\omega)r_\| - \omega t]\}\cdot\exp[-\alpha(\omega)r_\|]$, where $r_\|$ is the lateral coordinate.

Eqs.(12) and (13) yield two equations, namely, the dispersion relation

$$p(\omega) - f_1(\omega) = 0, \qquad (15)$$

And equation determining function $\alpha(\omega)$ or the propagation length $L(\omega) = \alpha^{-1}(\omega)$

$$\alpha(\omega) - f_2(\omega) = 0, \qquad (16)$$

where functions $f_1(\omega)$ and $f_2(\omega)$ must be clear from Eqs.(12) and (13).

In case of a pure real wavenumber $k'_{SP} \equiv p$ we have the complex frequency of the surface polariton

$$\omega_{SP}(p) = \omega'_{SP}(p) - i\omega''_{SP}(p) \equiv \omega'_{SP}(p) - i\Gamma(p), \qquad (17)$$

where $\Gamma(p)$ characterizes a temporal damping of the surface excitation.

In this case we obtain from Eqs.(7), (8) and (10) the equation of the fourth order with respect to the complex frequency of the surface polariton versus $p$

$$\omega_{SP}^4 + A\omega_{SP}^3 + B\omega_{SP}^2 + C\omega_{SP} + D = 0, \qquad (18)$$

where $A = i\nu$, $B = -(2p^2c^2 + \omega_P^2)$, $C = -i\nu 2p^2c^2$, $D = p^2c^2\omega_P^2$ for Eq.(7) and $A = i\gamma$, $B = -[p^2c^2(\varepsilon_\infty + 1) + \varepsilon_0\omega_{TO}^2]/\varepsilon_\infty$, $C = -i\gamma p^2c^2(\varepsilon_\infty + 1)/\varepsilon_\infty$, $D = p^2c^2\omega_{TO}^2(\varepsilon_0 + 1)/\varepsilon_\infty$ for Eq.(8).

The Eq.(18) yields two pairs of complex conjugate solutions. The required solution is chosen so that it does not include the growing factors and be appropriate for the propagation waves, for instance as $\exp\{i[pr_\parallel - \omega_{SP}(p)t]\} = \exp\{i[pr_\parallel - \omega'_{SP}(p)t]\} \cdot \exp[-\Gamma(p)t]$.

In case of the complex wavenumber and pure real frequency we have

$$D(\omega) = \frac{\sqrt{p^2(\omega) + \alpha^2(\omega)}}{2\pi} \sqrt{\left(\frac{dp(\omega)}{d\omega}\right)^2 + \left(\frac{d\alpha(\omega)}{d\omega}\right)^2}. \qquad (19)$$

It is not very difficult to verify that the real and imaginary parts of the wavenumber of a surface polariton can be decomposed within the frequency range $\omega \in [\omega_{TO}, \omega_{QP}]$ as follows

$$\begin{aligned} p(\gamma) &\approx a_1 + b_1\gamma^2 + c_1\gamma^4 + \dots, \\ \alpha(\gamma) &\approx a_2\gamma + b_2\gamma^3 + c_2\gamma^5 + \dots. \end{aligned} \qquad (20)$$

That is why we can neglecting by the imaginary $\alpha(\omega)$ part in Eq.(19) at the condition $\gamma \to 0$. As the next step instead of Eq.(19) we can write

$$D(\omega) = \frac{|p(\omega)|}{2\pi} \sqrt{\left(\frac{dp(\omega)}{d\omega}\right)^2 + \left(\frac{d\alpha(\omega)}{d\omega}\right)^2}, \qquad (21)$$

and then

$$D(\omega) \simeq \frac{|p(\omega)|}{2\pi}\left|\frac{dp(\omega)}{d\omega}\right|. \tag{22}$$

Using Eq. (8) and (12) in case of the transparent media ($\varepsilon'' \to 0$), we get from Eq.(22) the sought formula for the spectral DOS of surface polaritons

$$D(\omega) = \rho_0^{2D}\left\{\frac{\varepsilon'(\omega)}{\varepsilon'(\omega)+1} + \frac{\omega}{2[\varepsilon'(\omega)+1]^2}\frac{d\varepsilon'(\omega)}{d\omega}\right\}, \tag{23}$$

where $\rho_0^{2D}(\omega) = \omega/2\pi c^2$ is the two-dimensional (2D) spectral density of states with one polarization state.

Figure 1a) exemplifies the normalized DOS $\tilde{\rho}(\omega) = D(\omega)/D(\omega_{QP})$ of surface phonon polaritons supported by the plane vacuum-GaAs interface in accordance with Eq.(19) and Eqs.(12),(13). Normalization is done to $D(\omega_{QP})$ at the frequency of Coulomb surface polariton. The vertical dashed lines are situated at $\omega_{TO}$, $\omega_{QP}$ and $\omega_{LO}$ from the left to right side in the figures. We emphasize that the spectral validity of the results is limited by the condition $\omega_{TO} < \omega < \omega_{QP}$. For numerical calculations, we take the parameters corresponding to GaAs ($\omega_{TO} \approx 5.05\times 10^{13}\,rad/\sec$, $\omega_{LO} \approx 5.5\times 10^{13}\,rad/\sec$, $\gamma = 0.01\omega_{TO}$, $\varepsilon_\infty = 11$), from [28]. Figure 1b) shows normalized DOS of the surface phonon-polariton supported by the vacuum-GaAs interface versus a frequency calculated with use of Eq.(22) and Eq.(12). It is clearly seen two peaks in DOS. One larger peak is centered near the frequency of the quasistatic polariton $\omega_{QP}$. Additional smaller peak is connected with finite losses in our system due to an anharmonicity. We found that the smaller peak is centered near the frequency corresponding to the intersection point of the surface polariton wavelength and the propagation length.

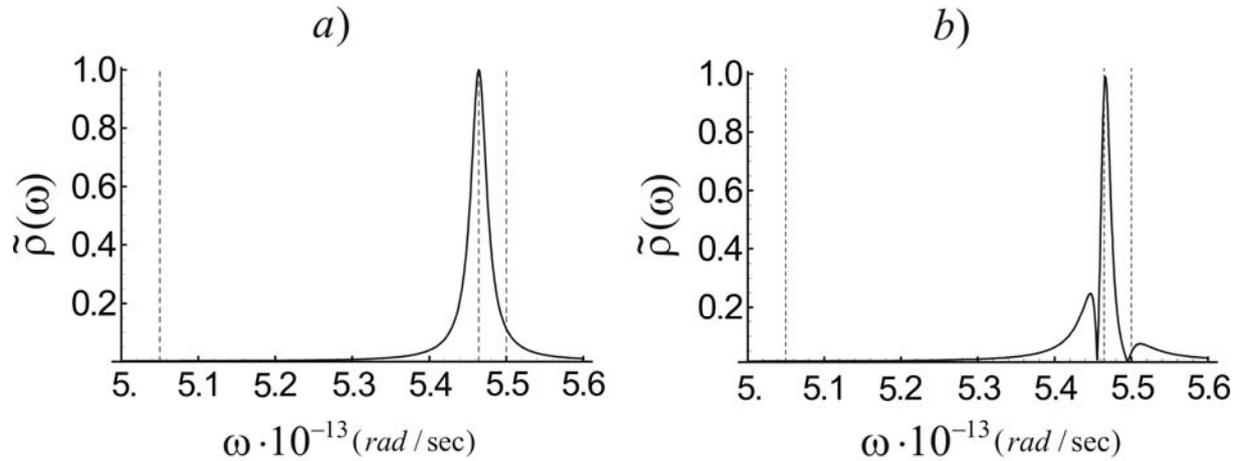

Fig.1

Figure 2 demonstrates normalized DOS $\tilde{\rho}(\omega) = D(\omega)/D(\omega_{QP})$ of the surface phonon-polariton supported by the vacuum-GaAs interface versus a normalized frequency $\omega/\omega_{QP}$ as calculated with use of Eq.(19) (upper curve), Eq.(21) (middle curve) and Eq.(22) (lower curve) at two different anharmonicities.

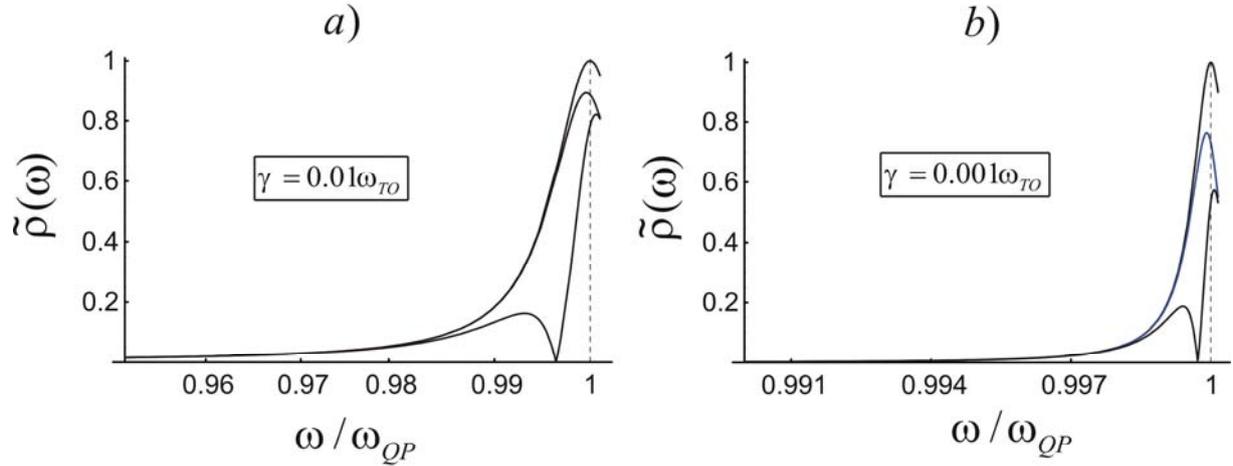

Fig.2

This figure shows how influence the terms $\alpha(\omega)$ and $d\alpha/d\omega$ in Eq.(19) determining a dissipation in a system on to the density of states.

Next figure demonstrates a connection between the smaller peak of DOS as calculated with help of Eq.(22) with a dissipation strength of material supported the surface polaritons.

Figure 3a) shows the normalized smaller peaks of DOS for GaAs as calculated with use of Eq.(22) at different anharmonicities $\gamma$. The vertical dashed line fixes a frequency of the quasistatic polariton $\omega_{QP} \approx 5.46 \times 10^{13} \, rad/\sec$ for this material at $\gamma = 0$. It is evidently that the smaller $\gamma$, the larger the peak of DOS and closer to $\omega_{QP}$. Figure 3b) shows the frequency positions of the peak maximum of DOS (thick line) from figure a) and of the intersection point (thin line) of the propagation length $L(\omega,\gamma) = \alpha^{-1}(\omega,\gamma)$ and wavelength $\lambda(\omega,\gamma) = 2\pi p^{-1}(\omega,\gamma)$ of the surface phonon-polariton at the GaAs/Vacuum interface calculated with help of the formulas Eqs.(12), (13), (14) at different anharmonicity $\gamma$. From this figure, we can conclude that the smaller peak of DOS as calculated with help of Eq.(22) is connected with dissipative properties of a system. The intersection point of $L$ and $\lambda$ determines two domains, where $L > \lambda$ and $L < \lambda$. In other words, the additional peak of DOS is determined by interplay between losses in a system and propagation conditions for surface polaritons.

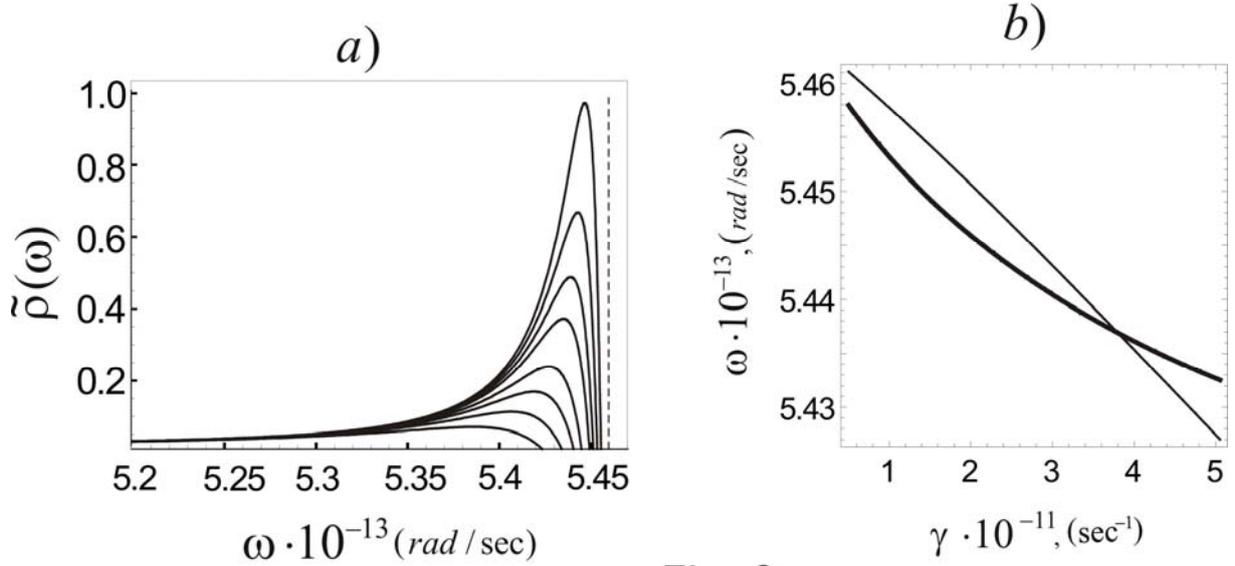

Fig.3

In the limit of $\gamma \to 0$ the additional peak merges with a peak of the normal mode at $\omega \simeq \omega_{QP}$. The same regularity was found also for the Drude model of dielectric function in [22]. We would like to emphasize that the finite dissipation rate in a system under study determine concrete and finite value of DOS and backbending in dispersion curve for surface polaritons [29-31].

Knowledge of DOS allows defining the spectral energy density of surface polaritons [21, 22] by analogy with the Planck's law for bulk photons as follows

$$u_{SP}(\omega) = D(\omega)\Theta(\omega,T), \qquad (24)$$

and the total energy density of surface polaritons

$$U_{SP}(T) = \int_{\omega 1}^{\omega 2} d\omega\, D(\omega)\Theta(\omega,T), \qquad (25)$$

where $\omega 1$ and $\omega 2$ determine the frequency range, where $\mathrm{Re}\{\varepsilon(\omega)\} < -1$. It is obviously, that other thermodynamic functions can be written by analogy with Eq.(24), (25). For example, for the free energy of the surface polaritons

$$F_{SP}(T) = -k_B T \int_{\omega 1}^{\omega 2} D(\omega)\, \ell n\left[Z(\omega,T)\right] d\omega, \qquad (26)$$

where

$$Z(\omega,T) = \left[Exp(\hbar\omega/2k_B T) - Exp(-\hbar\omega/2k_B T)\right]^{-1} \qquad (27)$$

is the partition function.

We would like to recall that the above described method of the DOS calculating is valid at very weak dissipative processes within a system under study.

It is known that the thermodynamic functions of a system in general include, aside of the volume contributions, other terms relating to the interfaces existing among of materials. Let's consider a subsystem of normal surface waves made of a plane boundary between the dissipative dispersive medium and vacuum.

### 3. Plane interface systems

*3.1. Function $\rho(\omega)$ for surface polaritons in local approach*

The normal modes of plane surface are the surface polaritons in the frequency range when the dielectric functions across the interface have opposite signs [23-27]. Because we know the dispersion relation for the surface modes, corresponding term to the free energy or energy from Eqs.(1),(2) may be clearly found. Indeed, in this case we have in Eq.(5) that $d\beta = k_\parallel dk_\parallel d\phi$, $\rho(\beta) = L^2/(2\pi)^2$ and in a local approach

$$D^{SP} = k_\parallel^2 - \frac{\omega^2}{c^2}\frac{\varepsilon(\omega)}{\varepsilon(\omega)+1}, \quad \mathrm{Re}\{\varepsilon(\omega)\} < -1, \qquad (28)$$

that is why it follows from Eq.(5) that the sought for quantity per unit area

$$\rho_{SP}(\omega) = -\frac{1}{2\pi^2}\int_0^\infty k_\parallel dk_\parallel \,\mathrm{Im}\left\{\left[D^{SP}\right]^{-1}\frac{\partial D^{SP}}{\partial \omega}\right\}, \qquad (29)$$

where

$$\frac{\partial D^{SP}}{\partial \omega} = -\frac{2\omega}{c^2}\left[\frac{\varepsilon(\omega)}{\varepsilon(\omega)+1} + \frac{\omega}{2[\varepsilon(\omega)+1]^2}\frac{d\varepsilon(\omega)}{d\omega}\right]. \qquad (30)$$

It should be noted that the equation for the free energy in Eq.(1) with Eq.(5) is valid for any dissipation strength. But, the quantity in Eq.(5) has no clear sense of the density of states in general [17,18], because this quantity may be of different sign for different frequency range. Nevertheless, in the limiting case of the transparent media, it is definitely the density of surface states within the specified frequency range. Namely, let's turn to the transparent media ($\mathrm{Im}\{\varepsilon(\omega)\} \to 0$) in Eq.(28). In this case we have from Eq.(28)

$$D^{SP} = k_\parallel^2 - \frac{\omega^2}{c^2}\frac{\varepsilon(\omega)}{\varepsilon(\omega)+1} \approx k_\parallel^2 - \frac{\omega^2}{c^2}\frac{\varepsilon'(\omega)}{\varepsilon'(\omega)+1} - i\Delta, \quad (\Delta \to 0), \qquad (31)$$

где $\Delta = (\omega^2/c^2)\varepsilon''(\omega)/[\varepsilon'(\omega)+1]^2$, where $\varepsilon'(\omega)$ and $\varepsilon''(\omega)$ are the real and imaginary parts of the dielectric function.

Applying Sohotskii's formula $\lim_{\Delta \to 0}(x - i\Delta)^{-1} = P(1/x) + i\pi\delta(x)$ in Eq.(29), we obtain the expression for the density of states of surface polaritons

$$\rho_{SP}(\omega) = \frac{\omega}{2\pi c^2} \left\{ \frac{\varepsilon'(\omega)}{\varepsilon'(\omega)+1} + \frac{\omega}{2[\varepsilon'(\omega)+1]^2} \frac{d\varepsilon'(\omega)}{d\omega} \right\}, \qquad (32)$$

identically to the result in [21,22] obtained from the wellknown definition of density of states usually accepted in theory of solids.

Another example we would like to refer to is the case of a plane-parallel film [25]. In a quasistatic regime ($c \to \infty$) we have the following pair of dispersion relations for two modes $\omega_+(k_{SP})$ and $\omega_-(k_{SP})$ of surface polaritons in a thin film of thickness $\ell$

$$D^{(+)} = k_+ + (2/\ell) ArcTanh[\varepsilon(\omega_+)], \qquad (33)$$

$$D^{(-)} = k_- + (2/\ell) ArcCoth[\varepsilon(\omega_-)]. \qquad (34)$$

Taking into account Eqs. (33), (34) and the definition in Eq.(5) we have $\rho(\omega) = \rho^{(+)}(\omega) + \rho^{(-)}(\omega)$, where

$$\rho^{(+)}(\omega) = -\frac{1}{\pi^2 \ell} \int_0^\infty k_+ dk_+ \; Im \left\{ \frac{1}{D^{(+)}} \frac{d\varepsilon(\omega_+)/d\omega_+}{1-\varepsilon^2(\omega_+)} \right\}, \qquad (35)$$

$$\rho^{(-)}(\omega) = -\frac{1}{\pi^2 \ell} \int_0^\infty k_- dk_- \; Im \left\{ \frac{1}{D^{(-)}} \frac{d\varepsilon(\omega_-)/d\omega_-}{1-\varepsilon^2(\omega_-)} \right\}. \qquad (36)$$

By considering the same transition $Im\{\varepsilon(\omega)\} \to 0$ we have from Eqs.(33) and (34)

$$D^{(+)} = k_+ - (2/\ell) ArcTanh[-\varepsilon'(\omega_+)] + i\Delta, \qquad (37)$$

$$D^{(-)} = k_- - (2/\ell) ArcCoth[-\varepsilon'(\omega_-)] - i\Delta, \qquad (38)$$

where $\Delta = (2/\ell)|\varepsilon''/(\varepsilon'^2 - 1)|$. Thus, Eqs.(35) and (36) can be transformed as follows

$$\rho^{(+)}(\omega) = -\frac{1}{\pi^2 \ell} \frac{d\varepsilon'(\omega_+)/d\omega_+}{1-\varepsilon'^2(\omega_+)} \int_0^\infty k_+ \delta\{k_+ - (2/\ell)ArcTanh[-\varepsilon'(\omega_+)]\} dk_+, \qquad (39)$$

$$\rho^{(-)}(\omega) = -\frac{1}{\pi^2 \ell} \frac{d\varepsilon'(\omega_-)/d\omega_-}{\varepsilon'^2(\omega_-)-1} \int_0^\infty k_- \delta\{k_- - (2/\ell)ArcCoth[-\varepsilon'(\omega_-)]\} dk_-. \qquad (40)$$

The integrals in these equations are not equal zero in case of $(2/\ell)ArcTanh[-\varepsilon'(\omega_+)] > 0$ and $(2/\ell)ArcCoth[-\varepsilon'(\omega_-)] > 0$. It means that it must be $-1 < \varepsilon'(\omega_+) < 0$ in Eq.(39) and $\varepsilon'(\omega_-) < -1$ in Eq.(40), correspondingly. That is why we have from (39) and (40) the following densities of states in this case

$$\rho^{(+)}(\omega) = -\frac{2}{\pi \ell^2} ArcTanh[\varepsilon'(\omega)] \left| \frac{d\varepsilon'(\omega)/d\omega}{1-\varepsilon'^2(\omega)} \right|, \qquad (\omega > \omega_{QP}), \qquad (41)$$

$$\rho^{(-)}(\omega) = -\frac{2}{\pi \ell^2} ArcCoth[\varepsilon'(\omega)] \left| \frac{d\varepsilon'(\omega)/d\omega}{1-\varepsilon'^2(\omega)} \right|, \qquad (\omega < \omega_{QP}), \qquad (42)$$

identically to the result in [21], where $\omega_{QP}$ is the frequency of a quasistatic plasmon-polariton satisfying the equation $\varepsilon'(\omega) = -1$.

We numerically calculated the function in Eq.(5) for case of the surface-plasmon and surface-phonon polaritons supported by a smooth interface between a vacuum and a homogeneous isotropic medium using local and nonlocal models of dielectric functions. For good metals we used the Drude model in local approach with plasma frequency $\omega_P = 1.42 \times 10^{16}\, rad/s$ for different dissipative parameter values $\nu$ (the frequency of collisions). The contribution of bounded electrons was modeled by a constant $\varepsilon_b = 3.6$. A frequency of the quasistatic plasmon-polariton in case of an ideal ($\nu = 0$) metal is equal $\omega_{QP} = \omega_P / \sqrt{1 + \varepsilon_b} \approx 6.62 \times 10^{15}\, rad/s$. For dielectric media we employed the oscillatory model of a permittivity with the frequency of transverse optical phonon $\omega_{TO} = 3.77 \times 10^{13}\, rad/s$ and $\varepsilon_0 = 9.06$, $\varepsilon_\infty = 5.8$ at different anharmonicities $\gamma$. A frequency of the quasistatic phonon-polariton is equal $\omega_{QP} \approx 4.59 \times 10^{13}\, rad/s$.

Figure 4 demonstrates the frequency dependence of the function $\rho(\omega)$ in Eq.(29) with dielectric functions of different kinds in a local approach. Figure 4a) exemplifies the frequency dependence of the normalized function $\tilde{\rho}(\omega, \gamma) = \rho(\omega, \gamma) / \rho_{max}(\omega, \gamma)$ in Eq.(29) at different $\gamma$ in a local model of the dielectric function. The numbers near curves correspond to the anharmonicity factor $\gamma = 0.03\omega_{TO}$ −1, $\gamma = 0.02\omega_{TO}$ −2, $\gamma = 0.015\omega_{TO}$ −3 and $\gamma = 0.01\omega_{TO}$ −4. The vertical dashed line represents position of the quasistatic surface phonon-polariton $\omega_{QP} \approx 4.59 \times 10^{13}\, rad/s$ for the chosen parameters of the medium.

In Fig.4b) the same dependence $\tilde{\rho}(\omega, \nu) = \rho(\omega, \nu) / \rho_{max}(\omega, \nu)$ is shown, but for good conductors in Eq.(29) at different damping factor $\nu$ in the Drude model of the dielectric function. The numbers near curves correspond to the damping factor $\nu = 0.02\omega_P$ −1, $\nu = 0.01\omega_P$ −2, $\nu = 0.005\omega_P$ −3 and $\nu = 0.001\omega_P$ −4. The vertical dashed line represents position of the quasistatic surface plasmon-polariton $\omega_{QP}$ for the chosen parameters of the conductor.

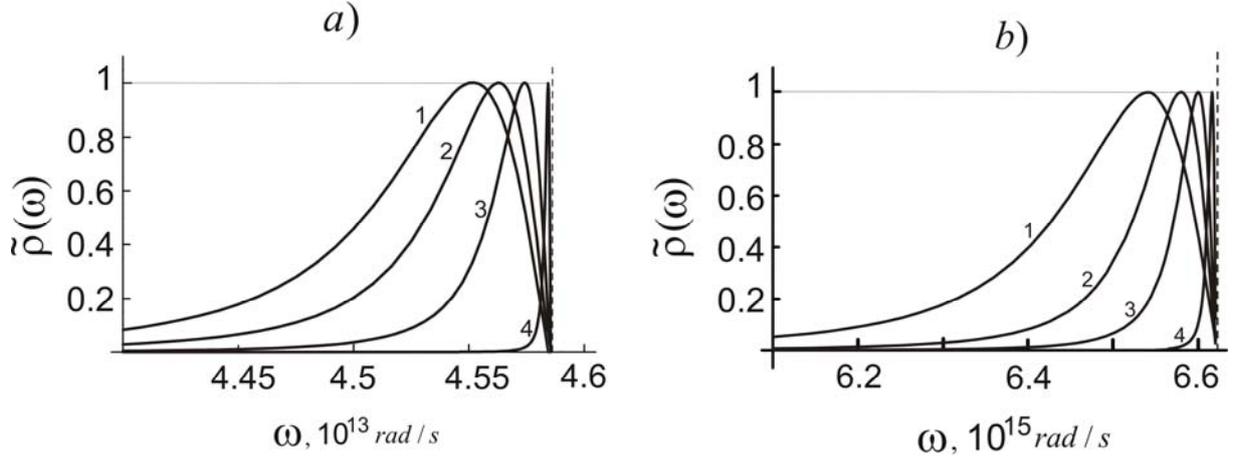

Fig.4

First of all, we would like to note that the results are qualitatively similar despite of the different models for dielectric functions. Both figures clearly show the peaked curves. The peaks are situated near the frequency of the quasistatic polaritons. As well as in our paper [22] we numerically show that the smaller the damping in a system, the closer these peaks to the frequency of the quasistatic surface polariton. The reason is the interplay between losses in a system and propagation conditions for surface polaritons. We especially emphasize that in a local approach the dispersion curves for surface polaritons are located within the "stop band" below the frequency $\omega_{QP}$ of the quasistatic polariton determined by the equation $\varepsilon(\omega) = -1$. In this approach the function $\rho(\omega) \geq 0$ in Eq.(29) and can be treated as the density of states.

In order to compare two definitions of DOS in Eq.(22) and in Eq.(29) within the frequency range $\omega \in [\omega_{TO}, \omega_{QP}]$, where the surface polaritons are supported, we show the normalized function $\tilde{\rho}(\omega, \gamma) = \rho(\omega, \gamma) / \rho_{max}(\omega, \gamma)$ at different $\gamma$ in Fig.5a) as calculated with help of Eq.(29) and the normalized smaller peaks of DOS as calculated with use of Eq.(22) also at different anharmonicities in Fig.5b) for GaAs. The smaller the anharmonicity, the larger the peak in these figures. We can to observe a qualitative agreement between the curves. Both definitions give the same shifts of peaks versus anharmonicity. Both definitions give similar qualitative dependence versus frequency within the frequency range, where $\text{Re}\{\varepsilon(\omega)\} < -1$. But in general, the usual definition of DOS gives the positively definite quantities in Eqs.(9),(21),(22) in accordance with the sense of the density of states. In its turn, the function $\rho(\omega)$ in Eq.(29) can not be interpreted as the DOS, because its negative values within some frequency intervals. We recall that in a nonlocal approach no the "stop band" [32] and within the band $\omega_{TO}(p) < \omega < \omega_{LO}(p)$ the propagating waves can exist despite of the sign of dielectric function.

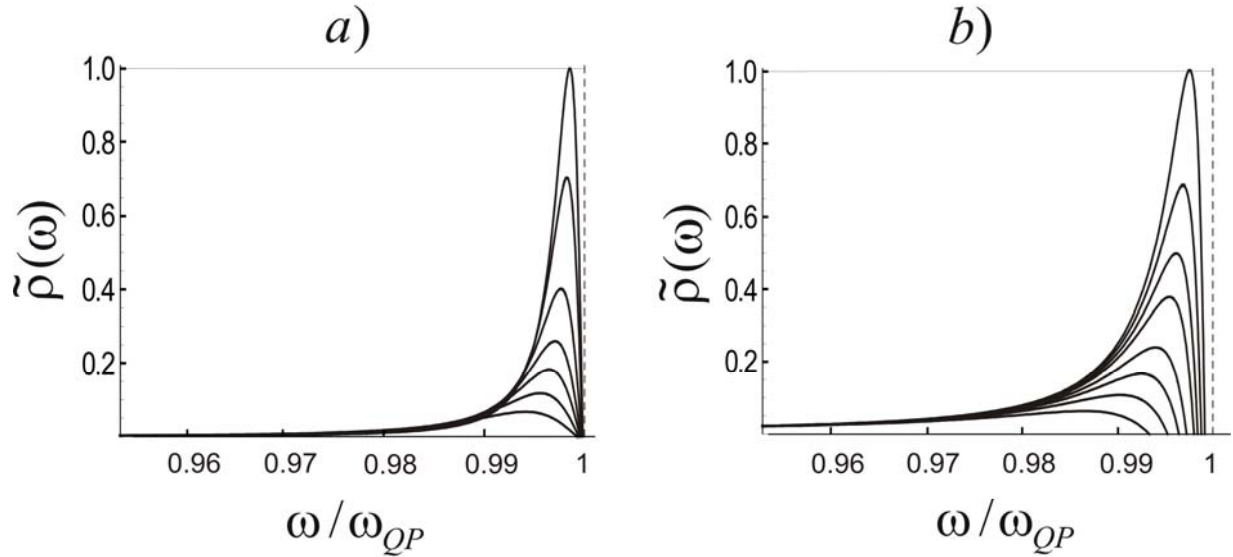

Fig.5

It should be emphasized that both of the definitions of DOS above described are identical in case of transparent materials. It is necessary to make one important remark concerning an integration in Eq.(5), which is performed over some volume $\Omega_\lambda$, separating the longwavelength part of the electromagnetic fluctuations [17,18]. It means that the integration with respect to the variable $k_\parallel$ in Eq.(29) is should be done up to some maximum value $\omega_P/c \ll p_{max} \ll a^{-1}$ for conductors, where $a$ is the interatomic distance, or $\omega_{TO}/c \ll p_{max} \ll a^{-1}$ for dielectrics, taking into account that $\varepsilon(k,\omega)$ tends to pure real value at $k_\parallel \to \infty$ in the nonlocal models of the dielectric functions in Eqs. (43) and (44). In this case the function $\rho(\omega) \to 0$ in Eq.(29). The exact magnitude of the upper limit of integration in local calculations with any given exactness can be established in comparison with corresponding calculations using the nonlocal models of dielectric functions. But, a calculation of the relative quantities is not very sensible to the choice of $p_{max}$.

*3.1. Function $\rho(\omega)$ for surface polaritons in non local approach*

In the presence of spatial dispersion, we took the modified oscillatory model [32] for nonconductors

$$\varepsilon(k,\omega) = \varepsilon_\infty + \frac{\Omega_P^2}{\omega_{TO}^2 - \omega^2 + D(k_\parallel^2 + q^2) - i\gamma\omega}, \tag{43}$$

where $\Omega_P^2 = (\varepsilon_0 - \varepsilon_\infty)\omega_{TO}^2$, $D = \hbar\omega_{TO}/m_{eff}$, $m_{eff} = 0.17 m_e$ for ZnSe, $k^2 = k_\parallel^2 + q^2$, and the hydrodynamic model for conductors [26] in a nonlocal case with the same plasma frequency, with the Fermi velocity $v_F = 1.4 \times 10^8 cm/s$

$$\varepsilon(k,\omega) = \varepsilon_B - \frac{\omega_P^2}{\omega^2 + i\nu\omega - \beta^2(k_\parallel^2 + q^2)}, \tag{44}$$

where $\beta^2 = (3/5)v_F^2$ for $\omega \gg \nu$.

Taking into account that the spatial structure of the surface fields is determined by the wave vector $\vec{k} = \{k_\parallel, q = i\alpha\}$, we found allowed values of $\alpha = \alpha(k_\parallel, \omega)$ associated with a given $k_\parallel$ and $\omega$ just as well as in [32]. From equations $(\omega/c)^2 \varepsilon(k,\omega) = k^2$, $\varepsilon(k,\omega) = 0$ and Eq.(21) it follows that

$$\alpha_{1,2}^2(k_\parallel, \omega) = (1/2)\left[ A \pm \sqrt{\tilde{A}^2 + 4\omega^2 \Omega_P^2/Dc^2} \right], \tag{45}$$

where $A = (k_\parallel^2 - \omega^2 \varepsilon_\infty/c^2) - \Gamma_B^2$, $\tilde{A} = (k_\parallel^2 - \omega^2 \varepsilon_\infty/c^2) + \Gamma_B^2$, $\Gamma_B^2 = (\omega^2 - \omega_{TO}^2 - Dk_\parallel^2 + i\gamma\omega)/D$ and $\alpha_3^2(k_\parallel, \omega) = \Omega_P^2/\varepsilon_\infty D - \Gamma_B^2$.

The three known roots for the hydrodynamic model in Eq.(44) are

$$\alpha_{1,2}^2(k_\parallel, \omega) = (1/2)\left[ B \pm \sqrt{\tilde{B}^2 + 4\omega^2 \omega_P^2/\beta^2 c^2} \right], \tag{46}$$

where in this case $B = (k_\parallel^2 - \omega^2 \varepsilon_B/c^2) - \Gamma_B^2$, $\tilde{B} = (k_\parallel^2 - \omega^2 \varepsilon_B/c^2) + \Gamma_B^2$, $\Gamma_B^2 = (\omega^2 - \beta^2 k_\parallel^2 + i\nu\omega)/\beta^2$ and $\alpha_3^2(k_\parallel, \omega) = \omega_P^2/\varepsilon_B \beta^2 - \Gamma_B^2$.

It should be noted that the permittivity in Eqs.(43) and (44) is a pure real quantity at $k \to \infty$ despite of a dissipation, thus, the function $\rho(\omega)$ in Eqs.(5) and (29) is equal to zero at the high limit of integration (for the shortwavelength range). The same statement is valid for the Lindhardt-Mermin dielectric function [33]. This remark should be kept in mind in integrating with respect to the wavenumber in Eqs.(5),(29) using the dielectric functions in a local approach. The mentioned models of dielectric functions in a local approach follow directly from Eqs.(43),(44) at $k \to 0$.

In a nonlocal approach no the "stop band" [32] because the dielectric function $\varepsilon(k,\omega)$ is the function of the wavenumber. It was shown in the cited work that, for instance, within the band $\omega_{TO}(k_\parallel) < \omega < \omega_{LO}(k_\parallel)$ the propagating waves can exist despite of the sign of dielectric function. Furthermore, no pure surface waves strongly localized near boundary, because the exponential tails $\exp[-\alpha_i(k_\parallel,\omega)z]$ within a matter are leaking to a medium due to a complexity of the factors $\alpha_i(k_\parallel,\omega)$, including a pure imaginary one in spite a frequency range. That is why taking into account a spatial dispersion the surface-like and volume-like modes are mixed inside such a material. As the result, the function $\rho(\omega)$ can accept both the positive and negative values within different frequency range.

Before calculating a frequency dependence of the function $\rho(\omega)$ for different materials, we would like to demonstrate the surface polaritons dispersion curves for given dielectric function in Eq.(43) in case of a weak spatial dispersion obtained without of the so-called additional boundary conditions. In this case [23] the dispersion relation has a usual form

$$k_\parallel^2 = \frac{\omega^2}{c^2} \frac{\varepsilon(\vec{k},\omega)}{\varepsilon(\vec{k},\omega)+1}, \qquad (47)$$

where $\vec{k} = \{k_\parallel, i\alpha_i\}$, $\alpha_i = \alpha_i(k_\parallel,\omega)$, $i$ means the $i$-th branch of waves within a nonlocal medium. In our case we use two kinds of waves, because the third branch gives identically $k_\parallel = 0$ due to equality $\varepsilon(k,\omega) = 0$ for $\alpha_3(k_\parallel,\omega)$. Taking into account Eq.(43) the wavenumber and frequency of the surface phonon-polariton are both complex, $k_\parallel = \text{Re}\{k_\parallel\} + i\,\text{Im}\{k_\parallel\}$ and $\omega = \text{Re}\{\omega\} + i\,\text{Im}\{\omega\}$. In such a case the dispersion curve depends on the fixed conditions of an experiment or calculation [23, 29-31]. Furthermore, it should be bear in mind that $\text{Re}\{\alpha_i(k_\parallel,\omega)\} > 0$, assigning the near surface modes.

As is known [21], dispersion of the volume polaritons is determined by the equation

$$\varepsilon(k_\parallel, q, \omega)\left[k_\parallel^2 + q^2 - \frac{\omega^2}{c^2}\varepsilon(k_\parallel, q, \omega)\right] = 0, \qquad (48)$$

where $k^2 = k_\parallel^2 + q^2$. In case of an isotropic and homogeneous medium we should take $k = k_\parallel$ and $q = 0$ in Eq.(48). It is easy to verify that the three dispersion equations directly follows from Eq.(48) using Eq.(43), namely $q_{1,2}^2 = -\alpha_{1,2}^2 = 0$ and $q_3^2 = -\alpha_3^2 = 0$. Thus, we have two possible functional dependencies; one of them $\omega(k_\parallel)$ gives three roots from equations

$$\omega_{1,2}^4 + (i\gamma)\omega_{1,2}^3 - \left[\frac{k_\parallel^2 c^2 + \Omega_P^2}{\varepsilon_\infty} + \omega_{TO}^2(k_\parallel)\right]\omega_{1,2}^2 - \left(i\gamma \frac{k_\parallel^2 c^2}{\varepsilon_\infty}\right)\omega_{1,2} + \frac{k_\parallel^2 c^2 \omega_{TO}^2(k_\parallel)}{\varepsilon_\infty} = 0, \quad (49)$$

$$\omega_3 = \sqrt{\omega_{LO}^2(k_\parallel) - i\gamma/2}, \quad (50)$$

where $\omega_{LO}^2(k_\parallel) = \omega_{TO}^2(k_\parallel) + \Omega_P^2/\varepsilon_\infty - \gamma^2/4$ and $\omega_{TO}^2(k_\parallel) = \omega_{TO}^2 + Dk_\parallel^2$. These equations determine complex frequencies $\omega_{1,2,3} = \text{Re}\{\omega\} + i\,\text{Im}\{\omega\}$ of three branches of the bulk polaritons as functions of the pure real wavenumber $k_\parallel$.

Another functional dependency $k_\parallel(\omega)$ gives three roots from equations

$$(k_\parallel)_{1,2}^4 + \left[\frac{\omega_{TO}^2 - \omega^2 - i\gamma\omega}{D} - \frac{\omega^2}{c^2}\varepsilon_\infty\right](k_\parallel)_{1,2}^2 - \left[\frac{\omega_{TO}^2 - \omega^2 - i\gamma\omega}{D} - \frac{\Omega_P^2}{\varepsilon_\infty D}\right]\frac{\omega^2}{c^2}\varepsilon_\infty = 0, \quad (51)$$

$$(k_\parallel)_3 = \sqrt{-\frac{\omega_{TO}^2 - \omega^2 - i\gamma\omega}{D} - \frac{\Omega_P^2}{\varepsilon_\infty D}}. \quad (52)$$

These equations determine complex wavenumbers $(k_\parallel)_{1,2,3} = \text{Re}\{k_\parallel\} + i\,\text{Im}\{k_\parallel\}$ of three branches of the bulk polaritons as functions of the pure real frequency $\omega$.

Obviously, from Eqs.(49)-(52) directly follow analytical formulas for various limiting cases, for instance, in case of a transparent medium, when $\gamma = 0.$ ,

In this our paper we use only Eqs.(49),(50) determining the real and imaginary parts of the bulk polariton's frequencies.

Figure 6 demonstrates the dispersion dependencies taking into account spatial dispersion for the surface polaritons from the implicit equation (47) numerically solved in form $\omega(k_\parallel)$ using Eq.(43) and for the bulk polaritons from explicit Eqs.(49),(50).

Figure 6a) exemplifies the wavenumber dependence of the real part of the frequency for the volume (V) and surface (S) polaritons in case of the weak spatial dispersion in according with Eqs.(47) and (48) using Eq.(43). Two Brewster's modes (B) are shown within a domain of $k_\parallel < \omega/c$, too. Dashed lines show the dispersion of the transverse $\omega_{TO}(k_\parallel)$ and longitudinal $\omega_{LO}(k_\parallel)$ optical phonons. Marked dashed lines 1,2,3 show lines $\omega = ck_\parallel$, $\omega = ck_\parallel/\sqrt{\varepsilon_\infty}$, $\omega = ck_\parallel/\sqrt{\varepsilon_0}$, correspondingly. Dotted line represents the dispersion of the quasistatic polariton in according with Eq.(53). The imaginary part of the surface and volume phonon-polaritons frequency is demonstrated in figure 6b) for the curves S1,2, V1 and B1,2 from figure 6a). As it seen from these figures the curves 1 and 2 intersect near the point ($\omega_{TO}$, $k_\parallel = \omega_{TO}/c$). In vicinity of the point the curves oscillate, and these jumps are artificially smoothed out.

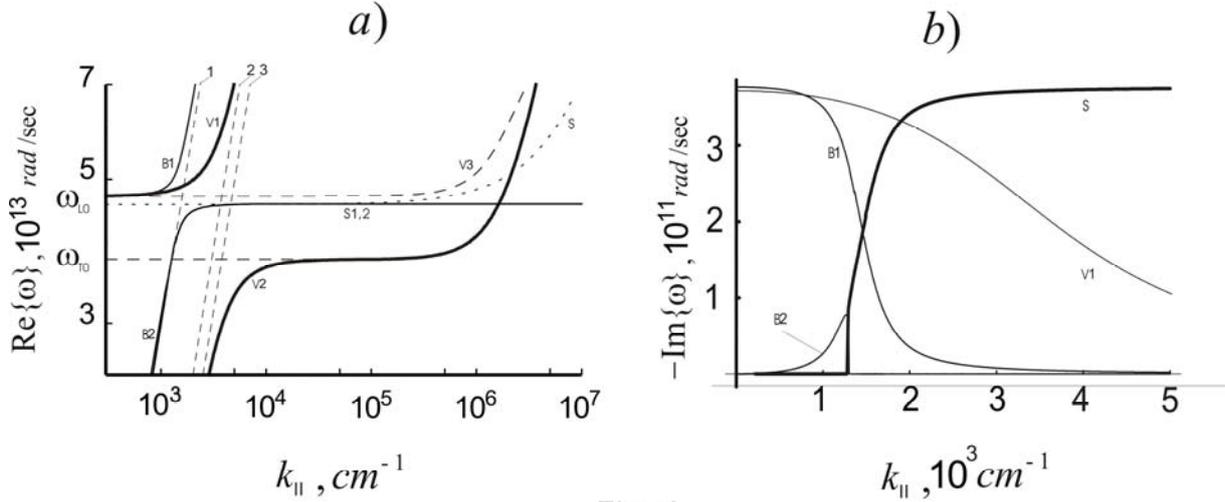

Fig.6

It should be noted from Fig.6a) that in a quasistatic limit ($k_\parallel \gg \omega/c$) a dispersion curve for the surface phonon-polariton S1,2 saturates due to Eq.(47) obtained without of additional boundary conditions [23]. There is more complicated dispersion equation, taking into account additional boundary conditions [32], and the wave number dependence of the quasistatic polariton in the above mentioned quasistatic limit

$$\omega_{QP}(k_\parallel) \simeq \omega_{QP} + \left(\frac{d\varepsilon}{d\omega}\bigg|_{\omega=\omega_{QP}}\right)^{-1}\left[\alpha_3\left(1+\frac{\Omega_P^2}{2D\Gamma_B^2}\right)+i\Gamma_B\right]\frac{k_\parallel}{i\alpha_3\Gamma_B}, \quad (53)$$

which is valid for the lossless ($\gamma = 0$) medium at the conditions $c \to \infty$ and $k_\parallel \to 0$. This dispersion is shown in Fig.6a) by the dotted line S.

Thus, we have dispersion equations (47),(48) for the surface and bulk waves of a semi-infinite medium described by the dielectric function (43) in a nonlocal approach to study the function $\rho(\omega)$ in Eq.(29) followed by calculation of thermodynamical properties of thermal fields within lossy and inhomogeneous materials.

Numerical calculations of the function $\rho(\omega)$ in Eq.(29) were done taking into account three roots $\alpha_{1,2,3}(k_\parallel,\omega)$. The three roots correspondingly yield three kinds of the dielectric functions in Eq.(43) or (44) and $D = D_1^{tr} D_2^{tr} D^\ell$ in Eq.(5). Three kinds of excitations give three contributions in the sought for function $\rho(\omega) = \rho_1^{tr}(\omega) + \rho_2^{tr}(\omega) + \rho^\ell(\omega)$. It is easily can be shown that the contribution of roots $\alpha_3(k_\parallel, \omega)$ in the accepted models of nonlocal permittivities to the total function $\rho(\omega)$ yields identically $\rho^\ell(\omega) = 0$, due to Eq.(48). The contributions from two roots in Eqs.(45),(46) corresponding to the transverse excitations was calculated as follows

$$\rho(\omega) = \rho_1^{tr}(\omega) + \rho_2^{tr}(\omega) = -\frac{1}{2\pi^2} \int_0^\infty k_\parallel dk_\parallel \, \text{Im}\left\{ \frac{\partial}{\partial \omega} \ell n\left[ D_1^{tr}(k,\omega) D_2^{tr}(k,\omega) \right] \right\}, \qquad (54)$$

where

$$D_i^{tr} = k_\parallel^2 - \frac{\omega^2}{c^2} \frac{\varepsilon_i(k,\omega)}{\varepsilon_i(k,\omega) + 1}, \qquad (55)$$

and

$$\frac{\partial D_i^{tr}(k,\omega)}{\partial \omega} = -\frac{2\omega}{c^2}\left[ \frac{\varepsilon_i(k,\omega)}{\varepsilon_i(k,\omega)+1} + \frac{\omega}{2[\varepsilon_i(k,\omega)+1]^2} \frac{d\varepsilon_i(k,\omega)}{d\omega} \right], \quad i=1,2, \qquad (56)$$

where $\varepsilon_i(k,\omega)$ is expressed by Eqs.(43) or (44) with $q = i\alpha_i(k_\parallel, \omega)$, $i = 1,2$ using Eqs.(45) or (46).

The results of our calculations are exemplified by figure 7. Figure 7a) demonstrates the frequency dependence of the normalized function $\tilde{\rho}(\omega) = \rho(\omega)/\rho_{\max}(\omega)$ in Eq.(54) at fixed anharmonicity $\gamma = 0.02\omega_{TO}$ in a nonlocal model of the dielectric function Eq.(43), where $\rho(\omega) = \rho_1^{tr}(\omega) + \rho_2^{tr}(\omega)$ corresponding to the two roots in Eq.(45). The vertical dashed lines represent positions of the transverse optical phonon frequency $\omega_{TO} \approx 3.77 \times 10^{13} \, rad/s$ (ZnSe, for example), the quasistatic surface phonon-polariton $\omega_{QP} \approx 4.59 \times 10^{13} \, rad/s$ and the longitudinal optical phonon frequency $\omega_{LO} \approx 4.71 \times 10^{13} \, rad/s$ at $k = 0$ and $\gamma = 0$ in Eq.(43). In Fig.7b) the frequency dependence of the function $\rho(\omega) = \rho_1^{tr}(\omega) + \rho_2^{tr}(\omega)$ in Eq.(54) corresponding to the two roots in Eq.(46) is shown at different dampings $\nu = 0.01\omega_P - 1$, $\nu = 0.007\omega_P - 2$, $\nu = 0.005\omega_P - 3$, for good conductors described by the nonlocal permittivity in Eq.(44). The vertical dashed line represents position of the quasistatic surface plasmon-polariton $\omega_{QP} = \omega_P / \sqrt{1+\varepsilon_b} \approx 6.62 \times 10^{15} \, rad/s$ at $k = 0$ and $\nu = 0$ in Eq.(44). The dimensionality of the function $\rho(\omega)$ here is equal $cm^{-2}(rad/s)^{-1}$.

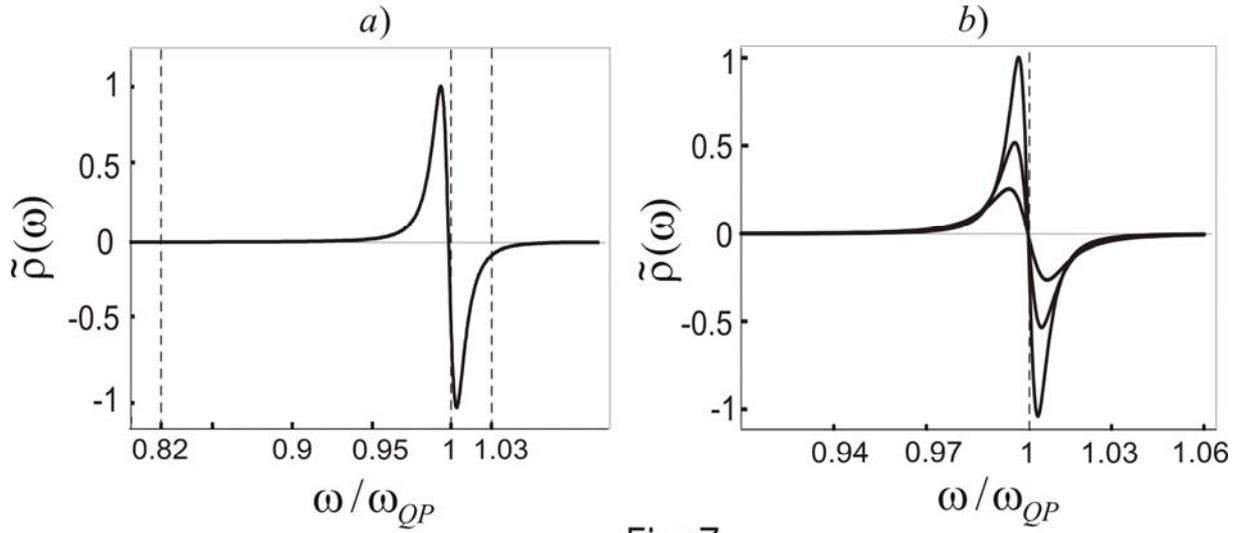

Fig.7

We reserved in Fig.7 the negative values of the function $\rho(\omega)$ within some frequency bands, because no strictly separated surface and bulk waves in general case of a lossy and nonlocal medium [32]. That is why taking into account a spatial dispersion the surface-like and volume-like modes are mixed inside such a material. In this case we should use all frequencies to calculate thermodynamic characteristics of surface polaritons.

In a comparative sense, we calculated the function $\rho(\omega)$ for the whole set of bulk and surface modes described by dispersion relations in Eqs.(47), (48) and exemplified by Fig.6.
Figure 8 shows the frequency dependence $\rho(\omega)$ for the bulk and surface modes; the curve 1 corresponds to the longitudinal wave $\rho^{\ell}(\omega)/(3\times 10^3)$ with a scale correction factor, the curve 2- to the two transverse waves $\rho_1^{tr}(\omega)+\rho_2^{tr}(\omega)$ as it is calculated for the bulk waves, and the curve 3 corresponds to $[\rho_1^{tr}(\omega)+\rho_2^{tr}(\omega)]\times 10^5$ with other correction factor for the surface waves. All calculations were done at fixed anharmonicity $\gamma = 0.02\omega_{TO}$. The vertical dashed lines represent positions of the transverse optical phonon frequency $\omega_{TO} \approx 3.77\times 10^{13}\, rad/s$, the quasistatic surface phonon-polariton $\omega_{QP} \approx 4.59\times 10^{13}\, rad/s$ and the longitudinal optical phonon frequency $\omega_{LO} \approx 4.71\times 10^{13}\, rad/s$ at $k=0$ and $\gamma = 0$ in Eq.(43). The dimensionality of the function $\rho(\omega)$ is equal $cm^{-r}(rad/s)^{-1}$, where $r=2,3$ for the surface or bulk states, correspondingly.

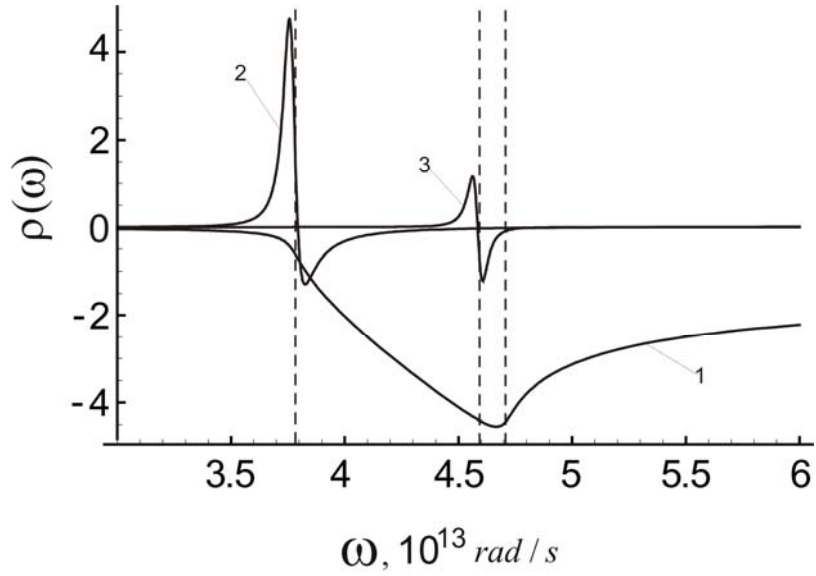

Fig.8

First of all, it should be emphasized that the longitudinal modes are exist only within a matter due to their fully electrostatic character. In this sense, they are bounded states manifesting in the negative sign of the function $\rho(\omega)$ in whole frequency range. Taking into account the correction factor for them in Fig.8, they are most energetic modes of thermal electromagnetic fluctuations. The function $\rho(\omega)$ for the transverse bulk and surface modes has contributions of both signs. From our point of view the negative tails can be attributed to the bounded waves in some frequency range, which are evanescent inside a matter. The evanescent character of these waves is totally due to the lossy mechanism within a matter. It is easy to verify that the negative part of the function $\rho(\omega)$ tends to zero when $\gamma \to 0$.

So far, we used the dispersion Eq. (10) in local approach or Eq. (47) for surface polaritons which is valid for very weak spatial dispersion. Now, we calculate the function $\rho(\omega)$ using more rigorous model accounting additional boundary conditions [29] as applied to the exciton resonance in the cubic semiconductor ZnSe. For convenience, we rewrite here the dispersion equation (4.12) from [29] as follows

$$D(k_\parallel, \omega) = (\alpha_1 + \alpha_0 \varepsilon_1)(i\Gamma_B + \alpha_1)(k_\parallel^2 - \alpha_3 \alpha_2) - $$
$$- (\alpha_2 + \alpha_0 \varepsilon_2)(i\Gamma_B + \alpha_2)(k_\parallel^2 - \alpha_3 \alpha_1) + \quad , \qquad (57)$$
$$+ k_\parallel^2 (i\Gamma_B + \alpha_3)(\alpha_2 - \alpha_1) = 0$$

where the dielectric functions $\varepsilon_{1,2}$ are defined in Eq.(43) with roots $\alpha_{1,2}$, correspondingly, $\alpha_0 = \sqrt{k_\parallel^2 - \omega^2 / c^2}$.

Using the general formula (54) with the dispersion equation (57) we numerically calculated the function $\rho(\omega)$ with chosen in [32] parameters. In our designations in Eq.(43) the background dielectric function $\varepsilon_\infty = 8.1$, $\omega_{ex} = 2.8\,eV \simeq 4.25 \times 10^{15}\,rad/s$ is equal $\omega_{TO}$ in Eq.(43), $\Omega_P = 3.15 \times 10^{14}\,rad/s$, $D = \hbar\omega_{ex}/m_e$ and with the parameter $\gamma \sim 10^{-5}\omega_{ex}$. All calculations were done at the conditions $\operatorname{Re}\{\alpha_{0,1,2,3}\} > 0$. In order to define the longwavelength electromagnetic fields, we integrated with respect to the value $k_\parallel$ from zero to some parameter $k_\parallel = a^{-1}$, determined the condition $k_\parallel a < 1$, where $a$ is the interatomic distance. Because we present relative function $\tilde{\rho}(\omega) = \rho(\omega)/\rho_{\max}(\omega)$ within the frequency range of our interest, the final result is not depend on the chosen parameter.

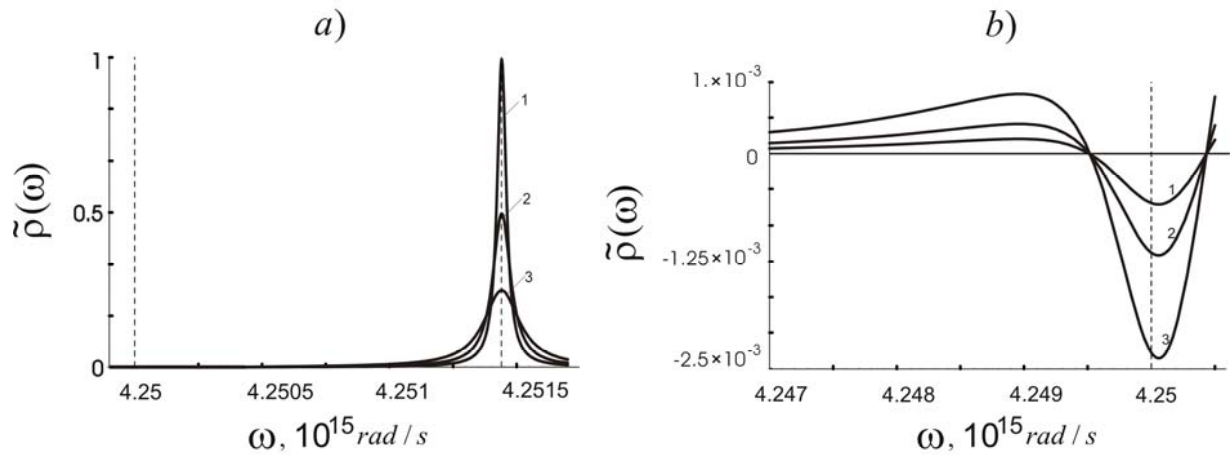

Fig.9

Figure 9 demonstrates the frequency dependence the normalized function $\tilde{\rho}(\omega) = \rho(\omega)/\rho_{\max}(\omega)$ as calculated with use of Eq.(54) with different parameter $\gamma = 10^{-5}\omega_{ex}$ - the curve 1, $\gamma = 2 \times 10^{-5}\omega_{ex}$-2, and $\gamma = 4 \times 10^{-5}\omega_{ex}$-3. Figure 9a) shows the positive part, and 9b) - the negative part of this function at the chosen parameters of $\gamma$. Relative contributions of two parts of this function depend on the dissipative parameter $\gamma$, and the negative part (in absolute values) can be larger than the positive one. As it seen from the figures the peaks are centered at frequencies $\omega_{ex}$ and $\omega = \sqrt{\omega_{ex}^2 + \Omega_P^2/\varepsilon_\infty}$. The negative peak appears with a certain dissipative parameter at its decreasing. It should be noted that the smaller the dissipative parameter $\gamma$, the smaller the negative peak up to disappearance at $\gamma \to 0$. We would like to note here that the negative part of $\rho(\omega)$ is essentially determined by dissipative mechanisms inside a medium.

## 4. Energy and free energy of the surface eigenmodes

In order to compare different definitions of DOS we numerically calculated the energy and free energy of surface phonon- and plasmon polaritons supported by a plane interface between a vacuum and a medium characterized by dielectric functions in Eqs.(7) and (8). Figures 10-13 show the normalized energy and free energy of the surface phonon- and plasmon polaritons as calculated with help of Eqs.(1),(25),(26) at different definitions of DOS in Eqs.(19),(21),(22),(29). The normalized energy $\tilde{U}(T) = U_{SP}(T)/U_{SP}(1000)$ of surface phonon polaritons supported by the plane vacuum-GaAs interface versus a temperature $T$ is demonstrated in Fig.10a) at different definitions of DOS in Eq.(29) – the curve 1, in Eqs.(19),(21) – the curves 2,3 and in Eq.(22) – the curve 4.

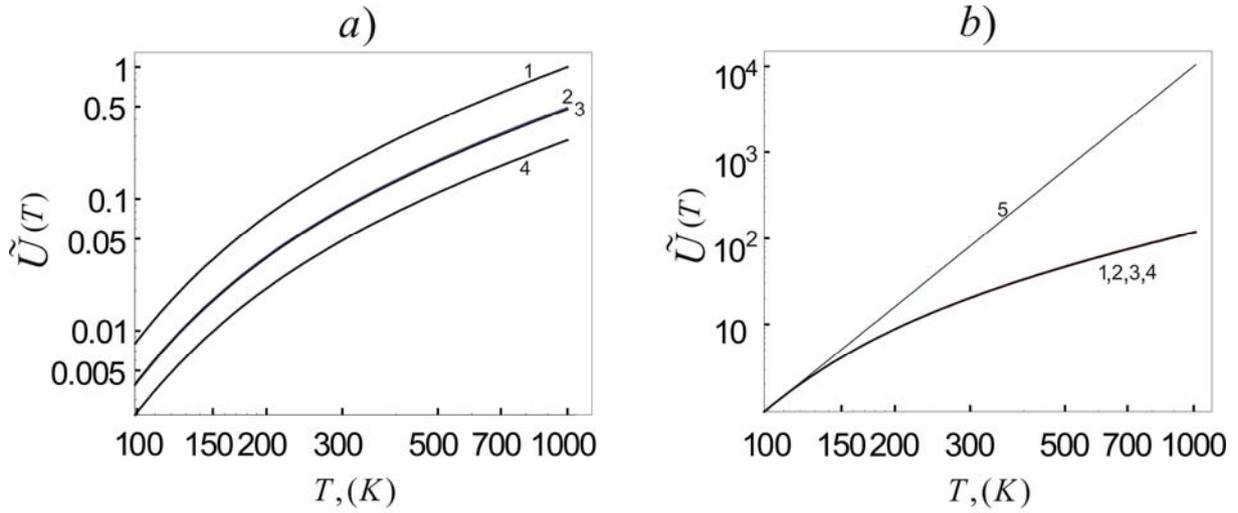

Fig.10

The same curves in comparison with the normalized energy of the black body photons in vacuum $U_{bb}(T) = \pi^2 (k_B T)^4 / 15(\hbar c)^3$ (the curve 5) is shown in Fig.10b). The normalization in this figure is done to $U_{bb}(100)$ for bulk photons and to $U_{SP}(100)$ for each curve, correspondingly. An integration in Eq.(25) is limited by the frequencies $\omega 1 = \omega_{TO}$ and $\omega 2 = \omega_{QP}$ taking into account the validity condition $\mathrm{Re}\{\varepsilon(\omega)\} < -1$ for surface phonon polaritons. It should be noted that the energy of surface phonon polaritons in Barash-Ginzburg theory

$$U_{SP}(T) = \int_{\omega 1}^{\omega 2} d\omega\, \rho(\omega)\Theta(\omega,T) = \int_{\omega 1}^{\omega 2} d\omega\, u_{SP}(\omega) \tag{58}$$

follows from Eq.(1) for free energy.

Figure 11 exemplifies the normalized energy of surface phonon polaritons supported by the plane interface "vacuum-SiC" versus a temperature at different definitions of DOS in Eqs.(19),(21),(22),(29). The normalizations are done as in Fig.10. The optical characteristics of $SiC$ in Eq.(7) are as follows $\omega_{TO} = 1.49 \times 10^{14}\, rad/s$, $\omega_{LO} = 1.82 \times 10^{14}\, rad/s$, $\gamma = 8.9 \times 10^{11}\, s^{-1}$, $\varepsilon_\infty = 6.7$ from [34], the frequency of the quasistatic phonon polariton $\omega_{QP} = 1.78 \times 10^{14}\, rad/s$.

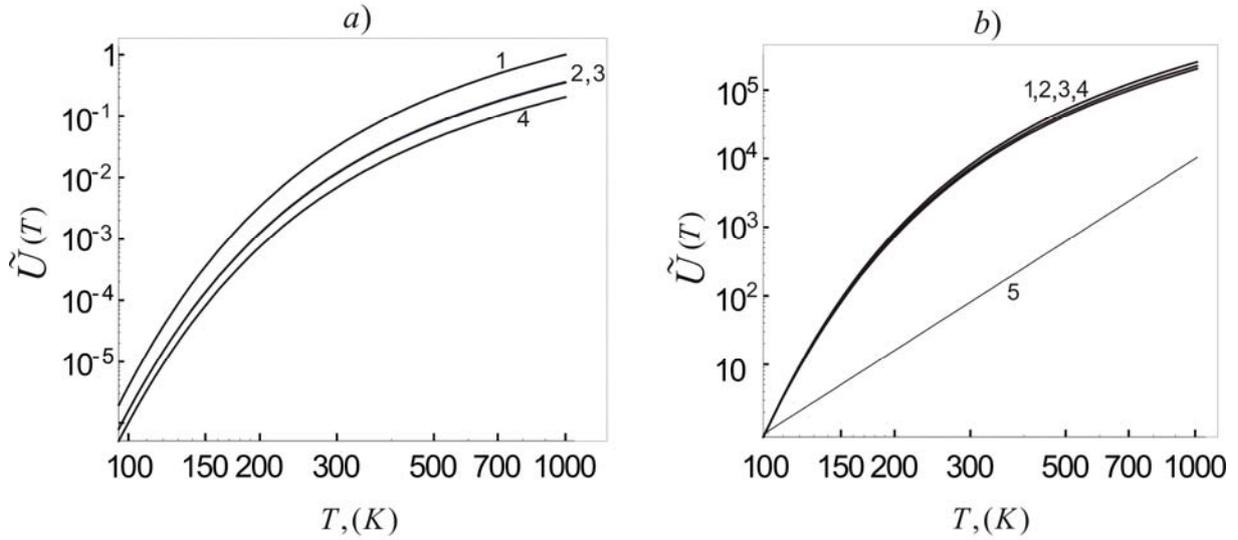

Fig.11

In calculations we used only the temperature dependent part of the mean energy of oscillator $\Theta(T) = \hbar\omega/[Exp(\hbar\omega/k_B T) - 1]$, omitting the zero-point term $\hbar\omega/2$.

It is seen from Figs.10, 11 that the energies have similar temperature dependence despite of different definitions of DOS. Moreover the temperature dependence of energy in case of surface phonon polaritons is quite different from the case of equilibrium bulk photons in vacuum.

The normalized free energies of the surface phonon polaritons versus a temperature as calculated with help of Eqs.(1),(26) at different definitions of DOS are represented in figures (12) and (13). Herewith, the normalized free energy $\tilde{F}(T) = \tilde{F}_{SP}(T)/\tilde{F}_{SP}(1000)$ of surface phonon polaritons supported by the plane vacuum-GaAs interface versus a temperature $T$ is demonstrated in Fig.12a) at different definitions of DOS in Eq.(29) – the curve 1, in Eqs.(19),(21) – the curves 2,3 and in Eq.(22) – the curve 4. The curve 5 everywhere in figures 12 and 13 shows the normalized free energy of the black body photons in vacuum $F_{bb}(T) = -\pi^2 (k_B T)^4 / 45(\hbar c)^3$. Fig.13a) shows the normalized free energy of surface phonon polaritons supported by the plane vacuum-SiC interface versus a temperature $T$ at different definitions of DOS.

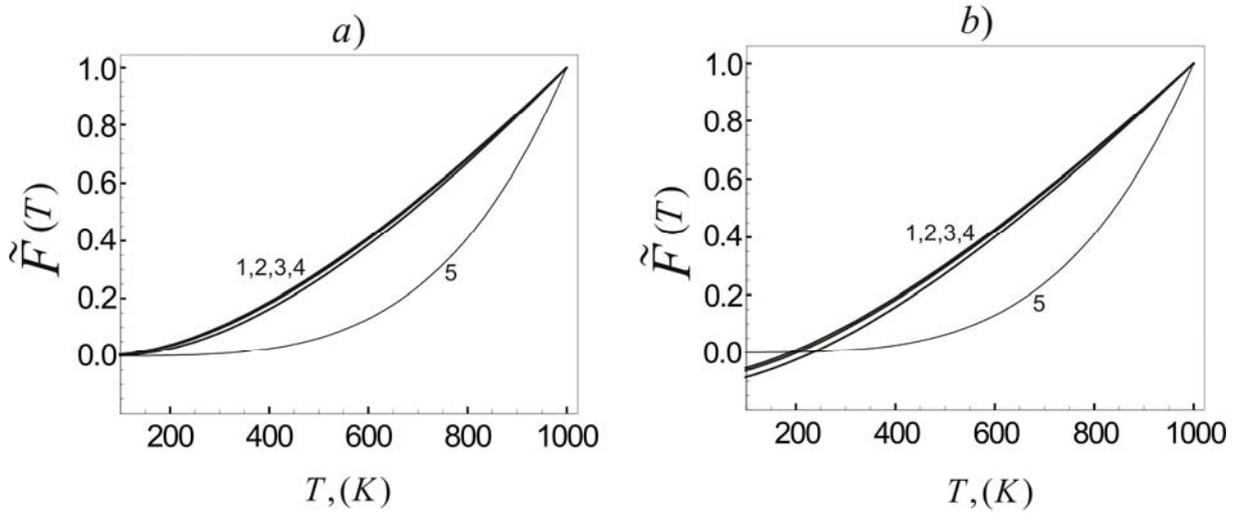

Fig.12

In figures 12a) and 13a) we used only temperature dependent parts of the free energy

$$F_{SP}(T) = k_B T \int_{\omega_{TO}}^{\omega_{QP}} D(\omega) \ln[1 - Exp(-\hbar\omega/k_B T)] d\omega, \qquad (59)$$

where $D(\omega)$ from Eqs.(19),(21) or (22),

$$F_{SP}(T) = k_B T \int_{\omega_{TO}}^{\omega_{QP}} \rho(\omega) \ln[1 - Exp(-\hbar\omega/k_B T)] d\omega, \qquad (60)$$

where $\rho(\omega)$ from Eq.(29).

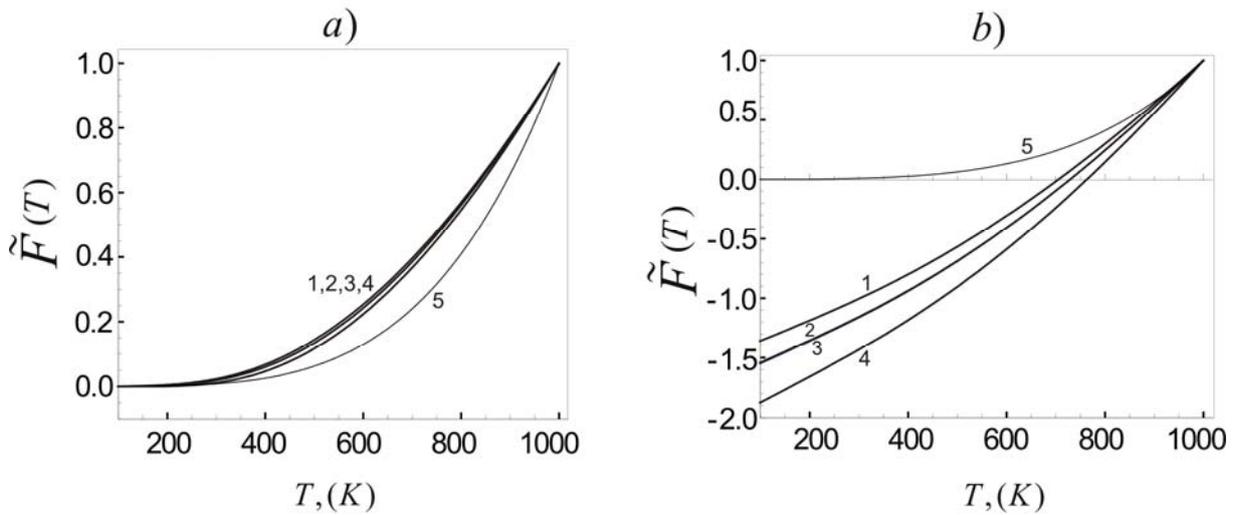

Fig.13

Figures 12b) and 13b) show the same dependencies, but including the zero-point term. In this case

$$F_{SP}(T) = k_B T \int_{\omega_{TO}}^{\omega_{QP}} D(\omega) \ell n[2 Sinh(\hbar \omega / 2 k_B T)] d\omega, \qquad (61)$$

and

$$F_{SP}(T) = k_B T \int_{\omega_{TO}}^{\omega_{QP}} \rho(\omega) \ell n[2 Sinh(\hbar \omega / 2 k_B T)] d\omega, \qquad (62)$$

with DOS $D(\omega)$ from Eqs.(19),(21) or (22) and $\rho(\omega)$ from Eq.(29), correspondingly.
Numerical calculations show that the discrepancy between of two definitions in Eq.(5) and Eq.(6) in case of good metals with finite dissipation strength is quite essential. Fig.14 demonstrates the normalized energy of surface plasmon polaritons supported by vacuum-aluminum interface.

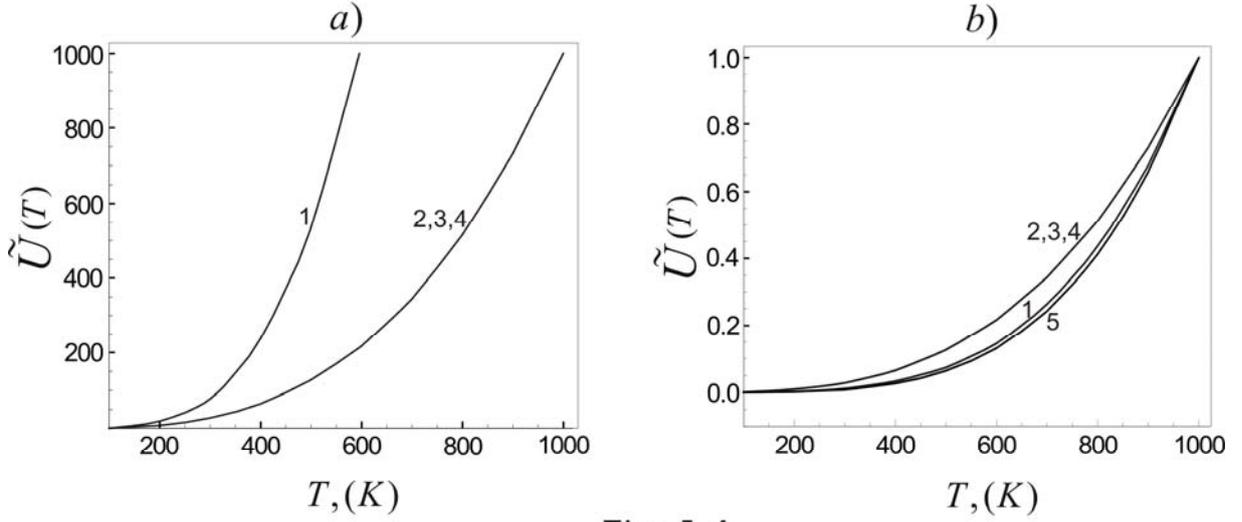

Fig.14

Herewith, the normalized energy $\tilde{U}(T) = \tilde{U}_{SP}(T) / U_{SP}(100)$ versus a temperature $T$ is shown in Fig.14a) at different definitions of DOS in Eq.(29) – the curve 1, in Eqs.(19),(21),(22) – the curves 2,3,4. In Fig.14b) the normalized energy $\tilde{U}(T) = \tilde{U}_{SP}(T) / U_{SP}(1000)$ is shown with other normalization factor. In this figure the normalized energy of equilibrium photons is also shown by the curve 5. Integration over a frequency range is done from $\omega 1 = 0$ up to $\omega 2 = \omega_{QP}$ in accordance with the validity condition $\text{Re}\{\varepsilon(\omega)\} < -1$ for surface plasmon polaritons.

As a final step, we compute the spectral power densities of surface plasmon polaritons at different DOS definitions to compare with the spectral power density of equilibrium photons in the Planck law $u_{bb}(\omega) = \hbar \omega^3 / \pi^2 c^3 [exp(\hbar \omega / k_B T) - 1]^{-1}$. Fig.15 shows the normalized spectral power density of equilibrium photons in accordance with the Planck law versus a frequency - the curve 1 and spectral power density of surface plasmon polaritons $\tilde{u}(\omega) = u_{SP}(\omega) / u_{SP}(\max)$ in accordance with Eq.(58) - the curve 2 and in accordance with Eqs.(9),(24)- the curve 3. Normalization is done to their maximum values in figure a) and to the maximum of the black body spectrum in figure b).

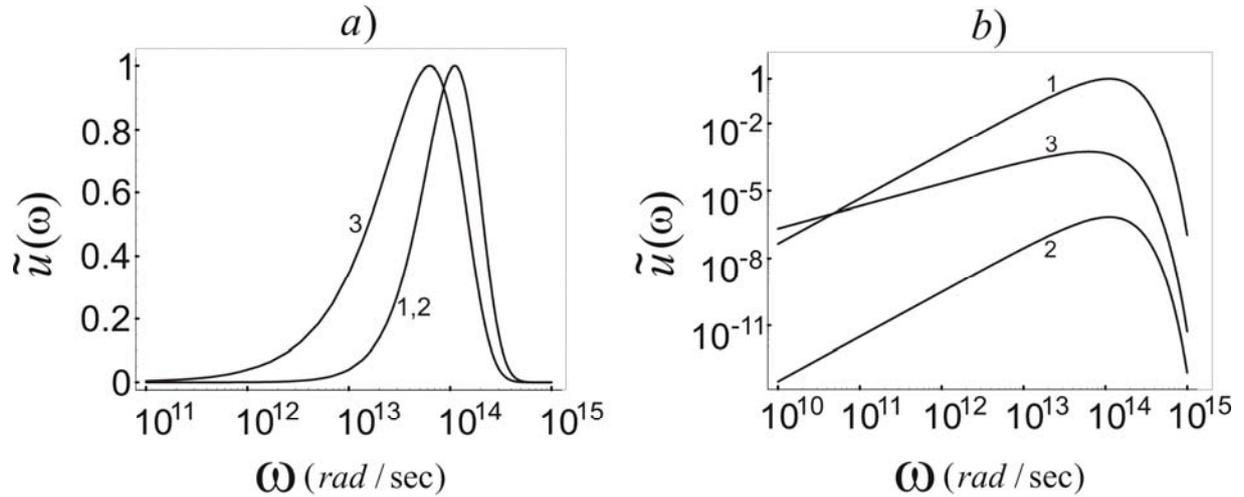

Fig.15

It is seen from these figures that the Barash-Ginzburg theory gives the same spectral distribution of surface plasmon polaritons as in the Planck law for bulk photons in a vacuum. In its turn, Eqs.(9),(24) yield in a shift of the spectral power density maximum towards lower frequencies with respect to the Planck law. Moreover, along with the quantitative difference in the spectral distribution, these spectral power densities are qualitatively different as it clearly seen from Fig.15. It should be emphasized that the Barash-Ginzburg theory gives DOS in Eq.(32) identically equal to Eq.(23) in case of transparent materials when $\nu \to 0$ and $\gamma \to 0$ in Eqs.(7),(8). But, the spectral shift in Fig.15a) exists despite of the dissipation factors in Eqs.(7),(8), that is why corresponding experiments would be very important.

## 5. Concluding remarks

Thus, from our calculations, see Fig.7, 8, 9 it follows that the negative tails of the function $\rho(\omega)$ in case of the transverse waves for 3D and 2D geometry are concerned with the dissipative conditions in a system of interest. In its turn, in case of the longitudinal waves this function is negative in all frequency range despite of a level of dissipation.

We recall that the function $\rho(\omega)$ in Eq.(1) cannot be treated as the density of states because of accepting the negative values in general. Taking into account that the integrand in Eq.(1) or in

Eq.(58) has a multiplicative form, we rewrite formally, for example Eq.(58), in another way as follows

$$U(T) = \int_0^\infty \rho(\omega)\Theta(\omega,T)d\omega \equiv \int_0^\infty D(\omega)E(\omega,T)d\omega, \qquad (63)$$

where $D(\omega)$ is explicitly the density of states, for instance the spectral density of states of free photons in vacuum with two independent polarizations

$$D(\omega) = \frac{\omega^2}{\pi^2 c^3}, \qquad (64)$$

or in two-dimensional case this is correspondingly the two-dimensional spectral density of states with one polarization state

$$D(\omega) = \frac{\omega}{2\pi c^2}, \qquad (65)$$

and the term

$$E(\omega,T) = \frac{\rho(\omega)}{D(\omega)}\Theta(\omega,T) = E^{(+)}(\omega,T) + E^{(-)}(\omega,T), \qquad (66)$$

which has a dimensionality of energy. This energy can be both of a positive $E^{(+)}(\omega,T)$ and negative value $E^{(-)}(\omega,T)$ in general. The negative values of energy correspond to bound states. In our case this is a subsystem of localized (evanescent) waves within a medium. Taking into account that the factor $\Theta(\omega,T)$ in the left integral of Eq.(58) is the mean energy of an isolated oscillator, the absolute value of the factor $|\rho(\omega)/D(\omega)|$ shows how many of the elementary oscillators are excited within a dissipative medium at the temperature $T$ due to fluctuations comparing with free space. The energy $E(\omega,T)$ can be treated as the field mean energy of the degree of freedom corresponding to the frequency $\omega$ within a dissipative medium at the temperature $T$. It should be noted that the field oscillators are attributed to the photons inside a matter for the transverse waves, but no the longitudinal photons in case of the longitudinal waves. Taking into account Eq.(63), an expression for the free energy

$$F(T) = -k_B T \int_0^\infty D(\omega) \ell n \left\{ [Z_0(\omega,T)]^{\frac{\rho(\omega)}{D(\omega)}} \right\} d\omega, \qquad (67)$$

and the Barash-Ginzburg formula (1) can be identically written as follows

$$F(T) = -k_B T \int_0^\infty D(\omega) \ell n [Z(\omega,T)] d\omega, \qquad (68)$$

where $Z(\omega,T)$ is the partition function of electromagnetic excitations within a material.

In conclusion, we summarize the results of our paper. In this paper we addressed to thermodynamic properties of surface electromagnetic waves supported by plane interfaces.

Spectral energy densities of surface plasmon and phonon polaritons and total specific energies in case of thermodynamic equilibrium are calculated using general definition of the DOS accepted in theory of solids and the Barash-Ginzburg theory for inhomogeneous and disspative media. It is demonstrated that the spectral distribution of surface polaritons in equilibrium, depending on the optical properties of the material, may be differ from the Planck's law for bulk photons. Temperature dependence of the surface energy density of the two dimensional electromagnetic fluctuations is differ from the temperature dependence of volume density of energy for photons in equilibrium is also shown. It is known that three dimensional fluctuations of fields in vacuum near a plane interface and two dimensional fluctuations of surface polaritons mainly occur at the frequency of the quasistatic surface polariton. But, a manifestation of corresponding resonances in the spectral energy density of thermal fluctuations depends on the exponential factor in the mean energy of an oscillator. Based on the Barash-Ginzburg general formulae for free energy of the electromagnetic fields within inhomogeneous and lossy materials, thermodynamic functions of the interface modes are computed using dispersion equations in various local and nonlocal approaches. In particular, the energy and free energy of thermally stimulated electromagnetic fields of surface eigenmodes of a plane interface and of a plane parallel film are considered. A correspondence between the Barash-Ginzburg approach in calculations of thermodynamic properties of normal modes within spatially nonuniform dispersive and dissipative systems and the traditional textbook definition of energy via the density of states of eigenmodes of spatially nonuniform dissipationless systems is established. It is shown that the Barash-Ginzburg basic definition of free energy of the surface electromagnetic fields is identical to the related expression as derived from the wellknown definition of density of states usually accepted in theory of solids in the case of dissipationless materials. Dispersion relations for the surface states in different approaches are considered and showed that negative values of the function $\rho(\omega)$ in the general Barash-Ginzburg theory can be concerned to the bounded states of fields within a material. The longitudinal bulk waves are bounded states manifesting in the negative sign of the function $\rho(\omega)$ in whole frequency range despite of dissipation. In its turn, the transverse fields inside bulk or surface systems contain only in part the coupled waves with a matter via some dissipative mechanism. It is follows that the Barash-Ginzburg theory gives qualitatively the same spectral distribution of surface plasmon polaritons as in the Planck law for bulk photons in a vacuum. In its turn, other considered approaches yield in a shift of the spectral power density maximum towards lower frequencies with respect to the Planck law. Moreover, along with the quantitative difference in the spectral distribution, these spectral power densities are qualitatively different as it clearly seen from demonstrated figures. This spectral shift exists despite of the dissipation factors in model dielectric functions that is why corresponding experiments would be very important.

**Acknowledgements**

This work was supported by Ministry of Education and Sciences of Nizhny Novgorod, the grant 11-02-97026.

**Figure captions**

**Fig.1.** Normalized DOS $\tilde{\rho}(\omega) = D(\omega)/D(\omega_{QP})$ of surface phonon polaritons supported by the plane vacuum-GaAs interface in accordance with Eq.(19) – a) and Eq.(22) – b). Normalization is done to $D(\omega_{QP})$ at the frequency of Coulomb surface polariton. The vertical dashed lines are situated at $\omega_{TO}$, $\omega_{QP}$ and $\omega_{LO}$ from the left to right side in the figures.

**Fig.2.** Normalized DOS $\tilde{\rho}(\omega) = D(\omega)/D(\omega_{QP})$ of the surface phonon-polariton supported by the vacuum-GaAs interface versus a normalized frequency $\omega/\omega_{QP}$ as calculated with use of Eq.(19) (upper curve), Eq.(21) and Eq.(22) (bottom curve) at two different anharmonicities: $\gamma = 0.01\omega_{TO}$ – a) and $\gamma = 0.001\omega_{TO}$ – b). Common normalization is done to $D(\omega_{QP})$ with use Eq.(19). The vertical dashed line is situated at $\omega_{QP}$.

**Fig.3.** Normalized smaller peaks of DOS for GaAs as calculated with use of Eq.(22) at different anharmonicities $\gamma = 0.005, 0.006, 0.007, 0.008, 0.01, 0.012, 0.015, 0.02\omega_{TO}$ from upper to bottom curves – a). The vertical dashed line is fixed at the frequency of quasistatic polariton $\omega_{QP} \approx 5.46 \times 10^{13} \, rad/sec$ for this material at $\gamma = 0$. Frequency positions of the peak maximum of DOS (thick line) from figure a) and of the point of intersection (thin line) of the propagation length $L(\omega, \gamma) = \alpha^{-1}(\omega, \gamma)$ and wavelength $\lambda(\omega, \gamma) = 2\pi p^{-1}(\omega, \gamma)$ of the surface phonon-polariton at the GaAs/Vacuum interface calculated with help of the formulas Eqs.(12), (13), (14) at different anharmonicity $\gamma$.

**Fig.4.** Frequency dependence of the normalized function $\tilde{\rho}(\omega, \gamma) = \rho(\omega, \gamma)/\rho_{max}(\omega, \gamma)$ in Eq.(29) at different anharmonicities $\gamma$ in the oscillatory model of the dielectric function - a). The numbers near curves correspond to the anharmonicity factor $\gamma = 0.03\omega_{TO} - 1$, $\gamma = 0.02\omega_{TO} - 2$, $\gamma = 0.015\omega_{TO} - 3$ and $\gamma = 0.01\omega_{TO} - 4$. The vertical dashed line represents position of the quasistatic surface phonon-polariton $\omega_{QP} \approx 4.59 \times 10^{13} \, rad/s$ for the chosen parameters of the medium. Frequency dependence of the normalized function $\tilde{\rho}(\omega, \nu) = \rho(\omega, \nu)/\rho_{max}(\omega, \nu)$ for good conductors in Eq.(29) at different damping factor $\nu$ in the Drude model of the dielectric function- b). The numbers near curves correspond to the damping factor $\nu = 0.02\omega_P - 1$, $\nu = 0.01\omega_P - 2$, $\nu = 0.005\omega_P - 3$ and $\nu = 0.001\omega_P - 4$. The vertical dashed line represents position of the quasistatic surface plasmon-polariton $\omega_{QP} = \omega_P/\sqrt{1+\varepsilon_b} \approx 6.62 \times 10^{15} \, rad/s$ for the chosen parameters of the conductor.

**Fig.5.** Normalized function $\tilde{\rho}(\omega, \gamma) = \rho(\omega, \gamma)/\rho_{max}(\omega, \gamma)$ of the surface polaritons supported by the vacuum/GaAs interface at different $\gamma$ in figure a) as calculated with help of Eq.(29) and the normalized smaller peaks of DOS as calculated with use of Eq.(22) also at different anharmonicities in figure b).

**Fig.6.** Dispersion curves for the volume (V) and surface (S) polaritons in case of the weak spatial dispersion in according with Eqs.(47) and (48) using Eqs.(43),(45). Two Brewster's modes (B) are shown within a domain of $k_{\parallel} < \omega/c$. Dashed lines show the dispersion of the transverse $\omega_{TO}(k_{\parallel})$ and longitudinal $\omega_{LO}(k_{\parallel})$ optical phonons in Eqs.(46). Marked dashed lines

1,2,3 show lines $\omega = ck_\parallel$, $\omega = ck_\parallel / \sqrt{\varepsilon_\infty}$, $\omega = ck_\parallel / \sqrt{\varepsilon_0}$, correspondingly. Dotted line represents the dispersion of the quasistatic polariton in according with Eq. (53).

The imaginary part of the surface and volume phonon-polaritons frequency is demonstrated in figure b) for the corresponding curves S1, 2, V1 and B1,2 in figure a).

**Fig.7.** Normalized function $\tilde{\rho}(\omega) = \rho(\omega)/\rho_{max}(\omega)$ versus the normalized frequency $\omega/\omega_{QP}$ as calculated with help of Eq.(54) and with use of the nonlocal models of the dielectric functions in Eq.(43) at $\gamma = 0.02\omega_{TO}$ – a) and in Eq.(44) – b). In figure b) this function is represented at different frequency collision rates $\nu = 0.01\omega_P, 0.07\omega_P, 0.05\omega_P$ from the bottom to upper curve. The vertical dashed lines are situated at frequencies $\omega_{TO}$, $\omega_{QP}$, $\omega_{LO}$ in figure a) and at $\omega_{QP} = \omega_P/\sqrt{2}$ in figure b).

**Fig.8.** Frequency dependence of the function $\rho(\omega)$ for the bulk and surface modes. The curve 1 corresponds to the longitudinal wave $\rho^\ell(\omega)/(3\times 10^3)$ with a scale correction, the curve 2 – to the two transverse waves $\rho_1^{tr}(\omega) + \rho_2^{tr}(\omega)$ as it is calculated, and the curve 3 corresponds to $[\rho_1^{tr}(\omega) + \rho_2^{tr}(\omega)] \times 10^5$ with indicated correction factor. All calculations were done at fixed anharmonicity $\gamma = 0.02\omega_{TO}$. The vertical dashed lines represent positions of the transverse optical phonon frequency $\omega_{TO} \approx 3.77\times 10^{13}\,rad/s$, the quasistatic surface phonon-polariton $\omega_{QP} \approx 4.59\times 10^{13}\,rad/s$ and the longitudinal optical phonon frequency $\omega_{LO} \approx 4.71\times 10^{13}\,rad/s$ at $k = 0$ and $\gamma = 0$ in Eq. (43). The dimensionality of the function $\rho(\omega)$ is equal $cm^{-r}(rad/s)^{-1}$, where $r = 2,3$ for the surface or bulk states, correspondingly.

**Fig.9.** Frequency dependence of the normalized function $\tilde{\rho}(\omega) = \rho(\omega)/\rho_{max}(\omega)$ as calculated with use of the dispersion equation (57) with different parameter $\gamma = 10^{-5}\omega_{ex}$ – the curve 1, $\gamma = 2\times 10^{-5}\omega_{ex}$ –2, and $\gamma = 4\times 10^{-5}\omega_{ex}$ –3. Figure 9a) shows the positive part, and 9b)– the negative part of this function at the chosen parameters of $\gamma$. The vertical dashed lines represent positions of the frequencies $\omega_{ex} = 4.25\times 10^{15}\,rad/s$ and $\omega = \sqrt{\omega_{ex}^2 + \Omega_P^2/\varepsilon_\infty} \approx 4.2511\times 10^{15}\,rad/s$.

**Fig.10.** Normalized energy $\tilde{U}(T) = U_{SP}(T)/U_{SP}(1000)$ of surface phonon polaritons supported by the plane vacuum-GaAs interface versus a temperature $T$ at different definitions of DOS in Eq.(29) – the curve 1, in Eqs.(19), (21) – the curves 2,3 and in Eq.(22) – the curve 4 in figure a). The same curves in comparison with the normalized energy of the black body photons in vacuum $U_{bb}(T)$ (the curve 5) in figure b). The normalization in figure b) is done to $U_{bb}(100)$ and to $U_{SP}(100)$ for each curve, correspondingly.

**Fig.11.** Normalized energy $\tilde{U}(T) = U_{SP}(T)/U_{SP}(1000)$ of surface phonon polaritons supported by the plane vacuum-SiC interface versus a temperature $T$ at different definitions of DOS in Eq.(29) – the curve 1, in Eqs.(19),(21) – the curves 2,3 and in Eq.(22) – the curve 4 in figure a). The same curves in comparison with the normalized energy of the black body photons in vacuum $U_{bb}(T)$ (the curve 5) in figure b). The normalizations are done as in Fig.10.

**Fig.12.** Normalized free energy $\tilde{F}(T) = \tilde{F}_{SP}(T)/\tilde{F}_{SP}(1000)$ of surface phonon polaritons supported by the plane vacuum-GaAs interface versus a temperature $T$ at different definitions of DOS in Eq.(29) – the curve 1, in Eqs.(19),(21) – the curves 2,3 and in Eq.(22) – the curve 4. The curve 5 in figures shows the normalized free energy of the black body photons in vacuum $F_{bb}(T)$. The curves are calculated with help of Eqs.(59),(60) in figure a) and with help of Eqs.(61),(62) in figure b).

**Fig.13.** Normalized free energy $\tilde{F}(T) = \tilde{F}_{SP}(T)/\tilde{F}_{SP}(1000)$ of surface phonon polaritons supported by the plane vacuum-SiC interface versus a temperature $T$ at different definitions of DOS as directly as in Fig.12.

**Fig.14.** Normalized energy $\tilde{U}(T) = \tilde{U}_{SP}(T)/U_{SP}(100)$ of surface plasmon polaritons supported by vacuum-aluminum interface versus a temperature $T$ at different definitions of DOS in Eq.(29) – the curve 1, in Eqs.(19),(21),(22) – the curves 2,3,4 in figure a). In figure b) the normalized energy $\tilde{U}(T) = \tilde{U}_{SP}(T)/U_{SP}(1000)$ with other normalization factor and the normalized energy of equilibrium photons (the curve 5) are shown.

**Fig.15.** Normalized spectral power density of equilibrium photons in accordance with the Planck law versus a frequency - the curve 1 and spectral power density of surface plasmon

polaritons in accordance with Eq.(58) - the curve 2 and Eqs.(9),(24) - the curve 3. Normalization is done to their own maximum values in figure a) and to the maximum of the Planck law in figure b).